\documentclass[a4paper,superscriptaddress,twocolumn,prb,aps,showpacs,floatfix,citeautoscript]{revtex4}
\usepackage{graphicx}
\usepackage{bm}
\usepackage{amsmath,amsbsy,amsfonts,mathrsfs}
\usepackage{longtable}
\usepackage{dcolumn}

\newcommand{\D}{{\rm d}}

\newcommand{\br}{{\bm r}}

\newcommand{\mint}[1]{\int\! \D^{3} #1 \, }
\newcommand{\mdint}[2]{\mint{#1}\!\!\!\mint{#2}}
\newcommand{\rdm}{\gamma}

\newcommand{\lyon}{Laboratoire de Physique de la Mati\`ere Condens\'ee et
  Nanostructures, Universit\'e Lyon I, CNRS, UMR 5586, Domaine scientifique de
  la Doua, F-69622 Villeurbanne Cedex, France}

\newcommand{\coimbra}{Centre for Computational Physics, Department of Physics, University of Coimbra, 3004-516 Coimbra, Portugal.}

\newcommand{\etsf}{European Theoretical Spectroscopy Facility.}

\newcommand{\tpci}{Theoretical and Physical Chemistry Institute, National Hellenic Research Foundation, Vas. Konstantinou 48, GR11635 Athens, Greece.} 

\newcommand{\berlin}{Institut f{\"u}r Theoretische Physik, Freie
    Universit{\"a}t Berlin, Arnimallee 14, D-14195 Berlin, Germany.}

\begin{document}

\title{Benchmark calculations for reduced density-matrix functional theory}

\date{\today}

\author{N.\,N. Lathiotakis}
\affiliation{\tpci}
\affiliation{\berlin}
\affiliation{\etsf}

\author{Miguel A.\,L. Marques}
\affiliation{\lyon}
\affiliation{\coimbra}
\affiliation{\etsf}

\begin{abstract}
  Reduced density-matrix functional theory (RDMFT) is a promising alternative approach
  to the problem of electron correlation. Like standard density functional
  theory, it contains an unknown exchange-correlation functional, for which
  several approximations have been proposed in the last years. In this article,
  we benchmark some of these functionals in an extended set of molecules with respect to
  total and atomization energies. Our
  results show that the most recent RDMFT functionals give very satisfactory results
  compared to more involved quantum chemistry and density functional 
  approaches.
\end{abstract}

\maketitle

\section{Introduction}

Reduced density-matrix functional theory (RDMFT) is based on Gilbert's 
theorem~\cite{gilbert} which guarantees that the expectation
value of any observable of a system in its ground-state is a unique functional
of the ground-state one-body reduced density-matrix (1-RDM). Thus, the
fundamental quantity in RDMFT is the 1-RDM instead of the electronic density
upon which DFT is built. Although this idea 
is relatively old, functionals of the 1-RDM have
been exploited for practical applications only in the last decade.

The total energy of the ground state of a system of $N$ electrons 
can be written in terms of the 1-RDM $\rdm$ as
\begin{multline}
  \label{eq:E_of_g}
  E_{\rm tot}\left[  \rdm \right] = 
  \mdint{r}{r'} \delta(\br - \br' ) \left[ -\frac{1}{2} \nabla_\br^
    2 \right] \rdm(\br, \br') \\
  +\mdint{r}{r'} \delta(\br - \br' ) v(\br) \rdm(\br, \br') \\
  +\frac{1}{2} \mdint{r}{r'} \frac{\rdm(\br,\br)\: \rdm(\br', \br')}{|\br-\br'|}
  + E_{\rm xc} \left[  \rdm \right] \,.
\end{multline}
The first term is the kinetic energy of the system. The second is the energy due
to an external potential $v(\br)$ acting on the system. This external potential,
which for simplicity is assumed to be local, 
is normally the ionic potential for atoms and molecules. The last two terms
account for the electron-electron interaction that can be cast into a usual Coulomb part 
and the remainder, which in the DFT fashion we will call exchange-correlation term $E_{\rm xc}$.
This last part is the only term that is unknown and, for 
practical applications of the theory, needs to be approximated. A great advantage
of RDMFT, compared to DFT, is that the kinetic energy term is an explicit functional
of the ground state $\rdm$. Another advantage is that $\rdm$ is a non-local quantity and 
contains more information than the electronic density.

The knowledge of the exact, or of a reasonable approximation, of the
functional~(\ref{eq:E_of_g}) allows for the minimization with respect
to $\rdm$ to determine the total energy and $\rdm$ for the ground
state. In basically all practical implementations of RDMFT, the
minimization with respect to $\rdm$ is replaced by a minimization with
respect to its eigenvalues $n_i$ and its eigenfunctions
$\varphi_i(\br)$, called respectively the occupation numbers and
natural orbitals.  In such a minimization, the N-representability
conditions for the 1-RDM~\cite{lowdin,coleman} have to be
enforced. These conditions are
\begin{equation}
  \sum_i^\infty n_i=N, \quad 0\leq n_i \leq 1
  \,.
\end{equation}
The first represents particle number conservation, while the
second reflects the Pauli antisymmetry principle. The first can be
enforced through the minimization of the quantity
\begin{equation}
  \label{eq:F}
  F=E_{\rm tot} - \mu \left( \sum_i^\infty n_i - N \right)
\end{equation}
instead of the total energy $E_{\rm tot}$. The quantity $\mu$ is the
appropriate Lagrange multiplier. A dramatic consequence of the second
condition is that it allows for occupation numbers being border
minima, i.e., exactly equal to one or zero. We refer to the
corresponding states as ``pinned states''. The fundamental difference
between these states and those with fractional occupancy is that for
the pinned states the derivative $\partial F / \partial n_p \neq 0$,
where $n_p$ is the occupation number of the pinned state. Thus,
$\delta F / \delta \rdm \neq 0$ at the minimizing $\rdm$.  It has been
demonstrated~\cite{nekjellium} that this is not an exceptional
situation but rather the rule for several approximate functionals of
the 1-RDM.

The $N$-representability conditions warrant that the trial $\rdm(\br,
\br')$ corresponds to either a pure $N$-electron state or to an ensemble
of $N$-electron states~\cite{coleman}. However, they do not warrant
that the reconstructed two-body reduced density-matrix (2-RDM) is also
$N$-representable.\cite{harriman} The $N$-representability of the
2-RDM is a much more complex problem.

Most of the approximations to $E_{\rm xc}$ do not depend explicitly on
$\gamma$, but are written in terms of the natural orbitals and the
occupation numbers. For this reason, RDMFT is also called natural
orbital functional theory. The first functional was devised by
M\"uller in 1984,~\cite{mueller} but it attracted little interest for
almost 15 years.  The situation changed due to the works of Goedecker
and Umrigar~\cite{UG} and Buijse and Baerends,~\cite{bb0} who
demonstrated that correlation energies for small atomic and molecular
systems calculated with the M\"uller functional (or simple
modifications of it) were in good agreement with the exact ones. Since
then, several promising new functionals have appeared in the
literature.~\cite{yasuda1,yasuda2,kollmar,kios1,kios2,pernal,csanyi,csgoe,gritsenko,piris,nekjellium}

The real quality of an approximate functional comes from its ability
to describe accurately electronic correlation and other properties of
real systems. Ideally, the performance of a functional should also be
uniform over a large class of systems. Furthermore, it is important to
know the limitations of existing functionals, as that increases their
reliability. In this respect, benchmarking has a fundamental role in
the development of any approach to electronic correlation.

Significant effort has been devoted to the assessment of the quality
of existing exchange-correlation functionals of the 1-RDM. A few of
the articles that proposed new functionals also presented example
calculations, normally for atoms or small molecular
systems.~\cite{UG,bb0,yasuda1,yasuda2,gritsenko,piris,leiva,piris_theo,piris_os,piris_lop,piris_JTCC,our_open_shell}
Particular attention has been payed to the performance of several
functionals for the whole dissociation curve of
dimers.~\cite{gritsenko,cohen,staroverov,herbert,piris_lop} Besides
atomic and molecular systems, RDMFT functionals have also been applied
to the homogeneous electron gas
(HEG).~\cite{ciospernal,csanyi,csgoe,nekjellium} This prototype
metallic system has proven quite useful in the evaluation and the
development of 1-RDM functionals.~\cite{csgoe,nekjellium} Apart from
the correlation energy, other properties were also calculated like
ionization potentials,~\cite{pernalip,leiva,piris_theo,piris_os} the
chemical hardness and the fundamental gap,~\cite{our_gap,piris_os}
dipole moments and static
polarizabilities,~\cite{pernal_barends,piris,piris_JTCC,piris_os} and
vibrational frequencies.~\cite{leiva} Finally, effort has been devoted
recently in the formulation of time dependent
RDMFT~\cite{TDRDMFT,TDRDMFT2}.

However, a thorough benchmark of the most used functionals for larger
molecular systems is still lacking. In this Article, we address this
issue, and present a systematic study of the most common
exchange-correlation functionals within RDMFT for the G2/97 molecule
set.~\cite{g2set1,g2set2} This set comprises 148 neutral molecules,
including 29 radicals, 35 non-hydrogen systems, 22 hydrocarbons, 47
substituted hydrocarbons, and 15 inorganic hydrides. Furthermore, the
molecules present in the G2/97 set are well known experimentally and
theoretically, which allows us to compare the relative merits of
current RDMFT functionals with other quantum chemistry approaches and
standard DFT. We present both correlation and atomization
energies.

The rest of this Article is structured as follows. In
section~\ref{sec:functionals}, we give an overview of the existing
1-RDM functionals with a special emphasis on those we chose to include
in our benchmark. Then, in section~\ref{sec:results}, we describe our
numerical method and discuss our results. Finally, in
section~\ref{sec:concl}, we conclude and give a brief outlook on the
present status of RDMFT.

\section{Functionals}
\label{sec:functionals}

In this section, we present some of the most known functionals of the
1-RDM that have been introduced to date.  A number of these, including
those of interest in the present work, can be cast into the form
\begin{multline}
  \label{eq:Exc}
  E_{\rm xc} = 
  -\frac{1}{2} \sum_{\stackrel{j,k}{\sigma}} 
  \mdint{r}{r'} f(n_{j\sigma}, n_{k\sigma}) 
  \\ \times
  \frac{\varphi_{j\sigma}^*(\br)\: \varphi_{k\sigma}^*(\br')\: \varphi_{k\sigma}(\br)\: 
  \varphi_{j\sigma}(\br')}{|\br-\br'|} \,,
\end{multline}
i.e., they have the form of the usual Hartree-Fock (HF) exchange
modified by the function $ f(n_{j\sigma}, n_{k\sigma})$ of the
occupation numbers. Functionals of this form are sometimes referred to
as of J-K type, as they only involve Coulomb (J) and exchange (K) type
integrals over the natural orbitals. The symbol $\sigma$ in
Eq.~(\ref{eq:Exc}) denotes the spin index and will be omitted in the
following, i.e., we restrict the presentation of functionals to closed
shell systems for simplicity. In that case, a factor of 2 should be
introduced in the expression~(\ref{eq:Exc}). We refer to
Refs.~\onlinecite{our_open_shell,piris_os} for the generalization of
the functionals to open shell systems.

The first approximation to $E_{\rm xc}$, introduced by M\"uller in
1984,~\cite{mueller} corresponds to the simple formula
\begin{equation}
  f^\text{M\"uller}(n_j, n_k) = \sqrt {n_j \: n_k}
  \,.
\end{equation}
In reality, M\"uller considered a more general form 
$f=n_j^\alpha \: n_k^{1-\alpha}$, but found that $\alpha=1/2$ 
was the optimal value. He
showed that the probability of finding an electron at $\br$ when a
second one is at $\br^\prime$ becomes negative in the neighborhood of
$\br^\prime$. This unphysical negative value is minimized for
$\alpha=1/2$. Buijse and Baerends~\cite{bb0} arrived at the same
formula for $f$ by modelling the exchange and correlation
hole. Interestingly, the reconstructed second density associated with
his functional satisfies the sum rule relating the second and the
first order densities. In addition, it yields the correct dissociation
limit of dimers of open shell atoms like H$_2$. Note that HF and also
standard DFT functionals fail for these cases. However, the M\"uller
functional overestimates substantially the correlation
energy~\cite{staroverov,herbert} of all systems it has been
applied to (including the HEG~\cite{ciospernal,csanyi,nekjellium}).
It was recently shown by Frank {\it et al.}~\cite{frank} that for two
electron systems the M\"uller functional provides a lower-bound for
the total energy. They also showed that, for this functional, the
total energy does not go to zero if the ionic potential is switched
off but to the value $-N/8$ a.u. They interpreted this value as an
effective self-interaction (SI) error and proposed a correction to the
M\"uller functional equal to $N/8$ a.u.

A modified form of the M\"uller functional was introduced
independently by Goedecker and Umrigar (GU).~\cite{UG} In the GU
functional the SI terms are removed from the xc term of the
Eq.~(\ref{eq:Exc}) and the direct Coulomb term. The corresponding
function $f$ takes the form
\begin{equation}
  f^\text{GU}(n_j, n_k) = \sqrt {n_j n_k} - \delta_{jk}
  \left(n_j - n_j^2 \right)
  \,.
\end{equation}
Contrary to the functional of M\"uller, GU is not consistent with the
sum rule relating the second and the first order densities.~\cite{UG}
Furthermore, it can not be expressed as an explicit functional of
$\rdm$.  However, it yields much better correlation energies than the
M\"uller functional for atoms and molecules at the equilibrium
geometry. Unfortunately, it fails to reproduce the correct
dissociation limit of small molecules.~\cite{staroverov,herbert}

Cs\'anyi and Arias,~\cite{csanyi} devised a functional by considering
a class of computationally feasible approximations of the two-body
density matrix as a finite sum of tensor products of single-particle
operators. They called their approximation corrected Hartree-Fock
(CHF). It reads
\begin{equation}
  f^{\rm CHF} = n_j n_k +
    \sqrt{n_j\left(1-n_j\right)}\sqrt{n_{k\phantom{j}\!\!\!}\left(1-n_k\right)}
  \,.
\end{equation}
They applied CHF and the M\"uller functionals to the
HEG,~\cite{csanyi} and found that the two functionals coincide in the
low density limit.  As the density increases, however, CHF
undercorrelates considerably and the solution collapses quickly to the
HF idempotent solution. The same is found for a series of two-electron
systems.~\cite{cohen,staroverov,herbert} For example, CHF gives the
idempotent solution for H$_2$ at equilibrium (but not for large
distances).~\cite{cohen,herbert} Thus, the tendency to fall to the
idempotent HF solution is a drawback of this functional, especially at
the equilibrium geometries. We also found this behavior for most of
the molecular systems we applied it to.  For this reason, we did not
consider this functional in our evaluation.

Noticing that, in the high density limit, CHF underestimates the
correlation energy while the M\"uller functional overcorrelates it
considerably, Cs\'anyi, Goedecker, and Arias (CGA)
introduced~\cite{csgoe} a form for the function $f$ which is an
average between these two functionals
\begin{equation}
  f^{\rm CGA} = \frac{1}{2}\left[n_j n_k +
    \sqrt{n_j\left(2-n_j\right)}\sqrt{n_{k\phantom{j}\!\!\!}\left(2-n_k\right)}\right]
  \,.
\end{equation}
This functional is very accurate in the high-density limit of the HEG,~\cite{csgoe}
and is comparable to common generalized-gradient approximations (GGA) for atoms
(albeit not as precise as the GU functional). Interestingly, it reproduces 
the correct dissociation limit for small molecules.~\cite{staroverov,herbert}

We now turn our attention to the most recent functionals of the 1-RDM.
A promising approach was recently put forth by Gritsenko, Pernal, and
Baerends.~\cite{gritsenko} They proposed three hierarchical
corrections in order to treat the overcorrelation of the M\"uller
functional: the BBC1, BBC2, and BBC3. For all these functionals it is
essential to divide the natural orbitals into strongly and weakly
occupied. This distinction appears naturally for finite systems since
usually in this case the natural orbitals have occupation numbers that
are either close to one or to zero. We denote the set of strongly
occupied orbitals as $S$ and the set of the weakly occupied as
$W$. For the BBC1, the function $f(n_j,n_k)$ is
\begin{equation}
  \label{eq:BBC1}
  f^{\rm BBC1}(n_j,n_k) = \left\{  \begin{array}{rl}
      -\sqrt{n_j\: n_k}\,, & j\neq k \wedge \phi_j, \phi_k\in W \,, \\[0.15cm]
      \sqrt{n_j\: n_k}\,,  & \mbox{\ otherwise,} 
    \end{array}
  \right.
\end{equation}
while for BBC2, we have
\begin{equation}
  \label{eq:BBC2}
  f^{\rm BBC2}(n_j,n_k) = \left\{  \begin{array}{rl}
      -\sqrt{n_j\: n_k}\,, & j\neq k \wedge \phi_j, \phi_k\in W\,, \\[0.15cm]
      n_j\: n_k\,,         & j\neq k \wedge \phi_j, \phi_k\in S\,, \\[0.15cm]
      \sqrt{n_j\: n_k}\,,  & \mbox{\ otherwise.} 
    \end{array}
  \right.
\end{equation}
The symbol $\wedge$ stands for the logical ``and'' while $\in$ means
that the orbitals on the left belong to the set on the right.  In
other words, the BBC1 functional consists in a sign change of the
function $f$ of the M\"uller functional for orbitals $i\neq j$ that
are both weakly occupied while the BBC2, in addition to BBC1
correction, restores the exchange-like form for orbitals $i\neq j$
that are both strongly occupied.  The reconstructed second order
density associated with BBC1 and BBC2 is also consistent with the sum
rule relating the second and first order densities.

Finally, in the BBC3 functional, there are corrections that apply only
to the bonding and anti-bonding orbitals. An issue that emerges
especially when BBC3 is applied to systems with large number of
electrons is the possibility of degenerate bonding and anti-bonding
orbitals. Selecting one of the degenerate orbitals as bonding or
anti-bonding breaks the symmetry of the molecule. For this reason, in
our treatment, we consider a modification that respects the
degeneracies of bonding and/or anti-bonding orbitals.  We subdivide
further the set $S$ into two subsets: the subset $S_b$ for the
degenerate bonding orbitals, and the subset $S_c$ for the rest.
Accordingly, we subdivide the set $W$ into the subset $W_a$
of the degenerate anti-bonding orbitals and $W_h$ of the rest weakly
occupied orbitals.  The function $f$ for the BBC3 becomes
\begin{widetext}
\begin{equation}
  \label{eq:BBC3}
  f^{\rm BBC3}(n_j,n_k) = \left\{ \begin{array}{rl}
      -\sqrt{n_j\: n_k}\,,  & \quad j\neq k\: \wedge \: \phi_j,\phi_k\in W\,, \\[0.15cm]
       n_j\: n_k\,,         & \quad j\neq k\: \wedge \: \phi_j,\phi_k\in S\,, \\[0.15cm]
       n_j\: n_k\,,         & \quad \left(\phi_j\in S_c \:\wedge\:\phi_k\in W_a\right) \vee
                                         \left(\phi_k\in S_c \:\wedge\:\phi_j\in W_a\right)\,, \\[0.15cm]
      n_j^2\,,              & \quad j= k\: \wedge \:\phi_j,\phi_k\in S_c\cup W_h\,, \\[0.15cm]
     \sqrt{n_j\: n_k}\,,    & \quad\mbox{\ otherwise,} 
    \end{array}
  \right.
\end{equation}
\end{widetext}
where $\cup$ stands for the union of two sets and $\vee$ for the
logical ``or''.  The first two cases in the definition~(\ref{eq:BBC3})
represents the BBC1 and BBC2 corrections, the third the inclusion of
the anti-bonding orbitals in the BBC2 correction (unless they interact
with bonding orbitals). Finally, the forth case stands for the removal
of the SI terms as in the GU functional for all orbitals apart from
those belonging to the bonding or anti-bonding subsets. This final
correction violates the sum rule between the second and first order
densities.

Gritsenko {\it et al.}~\cite{gritsenko} applied the BBC functionals to
diatomic molecules and showed that they give an accurate description
of these molecules at both the equilibrium distance and at the
dissociation limit.  BBC1 and BBC2 were also applied to the
HEG.\cite{nekjellium} BBC3 involves single orbital type of corrections
and for the HEG it coincides with BBC2.

Another functional that was recently introduced by
Piris~\cite{piris}, uses the cumulant expansion to derive a
reconstructed functional of the two-matrix.  Under further
considerations, Piris arrived at the functional (which we will call
PNOF0 from Piris Natural Orbital Functional)
\begin{equation}
  f^{\rm PNOF0} = \Lambda_{jk}^{(0)} - \delta_{jk}
  \left(n_j - n_j^2 \right)
\end{equation}
with
\begin{equation} 
  \Lambda_{jk}^{(0)} = [ 1-2\theta(\frac{1}{2} - n_j) \: \theta(\frac{1}{2} - n_k )]\sqrt{n_j\: n_k}\,,
\end{equation}
where $\theta$ is the Heaviside step function.  This approximation can
be seen as a combination of the removal of the SI terms, as in the GU
functional, applied on top of the BBC1 correction.  In order to avoid
the effect of state pinning, Piris modified $\Lambda_{jk}$ by adding
an extra term. We refer to this functional as PNOF. It reads
\begin{equation}
f^{\rm PNOF} = \Lambda_{jk} - \delta_{jk}
  \left(n_j - n_j^2 \right)\,, 
\end{equation}
with
\begin{multline} 
\Lambda_{jk} = \Lambda_{jk}^{(0)}
        + \theta(n_j - \frac{1}{2} ) \: \theta(n_k - \frac{1}{2} )\: \sqrt{(1-n_j)\: (1-n_k)}
\end{multline}

PNOF was applied to a set of 57 molecules by calculating the correlation
energies and dipole moments and comparing with coupled cluster theory
and experimental values.\cite{piris} It was also applied to the
calculation of ionization potentials and vibrational
frequencies\cite{leiva,piris_theo,piris_os}. Recently a generalization
of this functional to open shell systems was
introduced,\cite{piris_os} and applied to the calculation of
correlation energies, ionization potentials and electron
affinities. Finally, a new ansatz based on PNOF\cite{piris_lop} has
been introduced for the calculation of dispersion forces and applied
to the calculation of helium dimer.

\section{Results}
\label{sec:results}

\section{Implementation}

Our implementation of RDMFT for finite systems is based on an
expansion of the natural orbitals on Gaussian basis sets. The total
energy as well as its gradients with respect to the occupation numbers
and the natural orbitals can then be written in terms of the so-called
one and two-electron integrals. Thus, the numerical treatment can be
based on a HF implementation provided that the functional is of J-K
type. Note that all functionals we are concerned with in this Article
are of this type. The required one and two-electron integrals are
inputs to our program. In the current implementation, we have been
using the GAMESS program~\cite{gamess} for this task.

We minimize the total energy of Eq.~\eqref{eq:E_of_g} following
two steps that are repeated alternately. In the first, we minimize
with respect to the occupation numbers keeping the natural orbitals
fixed. For this step, we use a sequential quadratic programming
method.~\cite{fletcher} For the second, we use a conjugate gradient
method. We use appropriate energy gradients which respect the orbital
orthonormality.~\cite{goedecker,cohen} The gradients of the total
energy with respect to the natural orbitals have the form
\begin{equation}
  \frac{\delta E_{\rm tot}}{\delta \phi_i^*}=F^{(i)}\: \phi_i\,,
\end{equation} 
where $F^{(i)}$ is a generalization of the Fock matrix of HF theory,
that in RDMFT depends on the orbital $i$. Consequently, the problem
does not reduce to a simple diagonalization, and the matrices
$F^{(i)}$ have to be updated very often in the minimization process.
This turns out to be the most time-consuming part of our
implementation, since the calculation of $F^{(i)}$ involves the
summation over the two electron integrals. A possible workaround to
the orbital minimization problem has been proposed by
Pernal.~\cite{PernalOEP} In this approach, all natural orbitals stem
from the same non-local potential that is obtained from a procedure
inspired by the optimized effective potential method (OEP) of
DFT.

We used two sets of calculations: The first set includes all molecules
of the G2/97 set, calculated with the Cartesian 6-31G$^*$ Gaussian
basis set. In the second, we employed the significantly larger cc-pVDZ
basis set but we considered a subset of the G2/97 set (G2-1), that
includes 55 molecules. In this way, we could check the dependence of our
findings on the choice of basis set. As reference we used the coupled
cluster method with singles, doubles and perturbative triples
[CCSD(T)]~\cite{ccsdt} employing the same basis sets.

Apart from the orbitals with occupancy close to one, we included 40
extra natural orbitals in the minimization procedure.  It has been
shown that this number is sufficient to achieve convergence of the
total energy.~\cite{UG,our_open_shell} For systems that are small
enough and this number exceeds the size of the basis, we performed full
variation, i.e., the number of the varied natural orbitals was equal to
the size of the basis.  For comparison, we computed the correlation
energy using DFT with the Becke 3 parameter exchange-correlation
functional (B3LYP)~\cite{b3lyp1,b3lyp2} and M{\o}ller Plesset
second-order perturbation theory (MP2). For these, as well as for the
CCSD(T) calculations, we used the Gaussian 98
program.~\cite{gaussian98} In all calculations, the geometries
optimized with the MP2 method were used.

For the open shell systems, our implementation~\cite{our_open_shell}
assumes spin dependent occupation numbers but spin independent natural
orbitals. Furthermore, in the minimization with respect to the spin
dependent occupation numbers, both spin-up and spin-down total numbers
of electrons were kept fixed. One can show that in this formulation the
$S_z$ component of the total spin of the system is
preserved. Moreover, this approach is very efficient numerically, as
the number of orbitals included in the variation process is half the
number for the full unrestricted case. We should mention that there is
not a unique way to generalize 1-RDM functionals to the case of
open shell systems. For example, a generalization, alternative to the
one used in this work, was recently introduced for the PNOF
functional.\cite{piris_os}

\section{Total energy}

\begin{figure}
  \centering
     \begin{tabular}{c}
        \includegraphics[width=7.5cm]{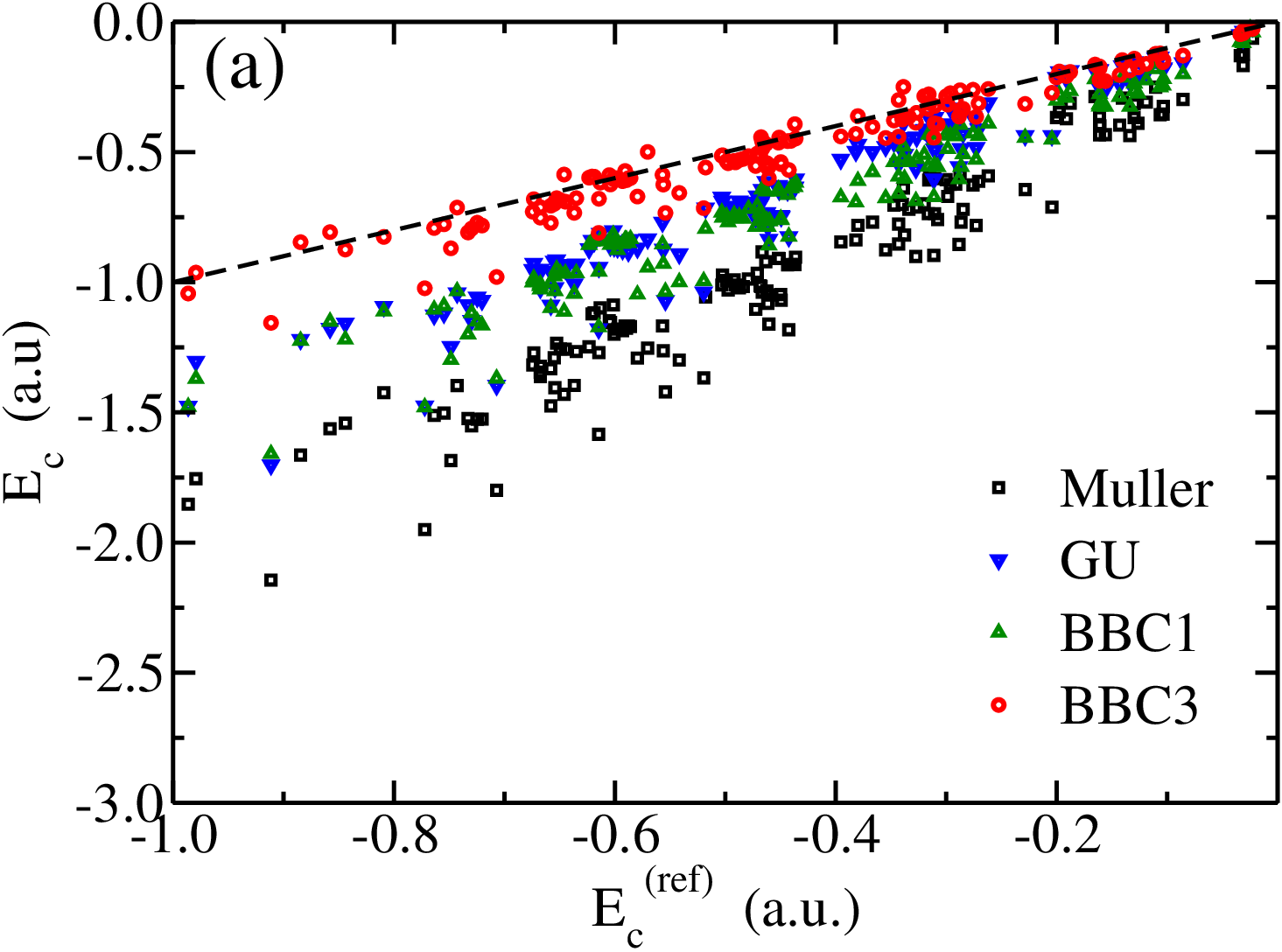} \\
          \\
          \\
        \includegraphics[width=7.5cm]{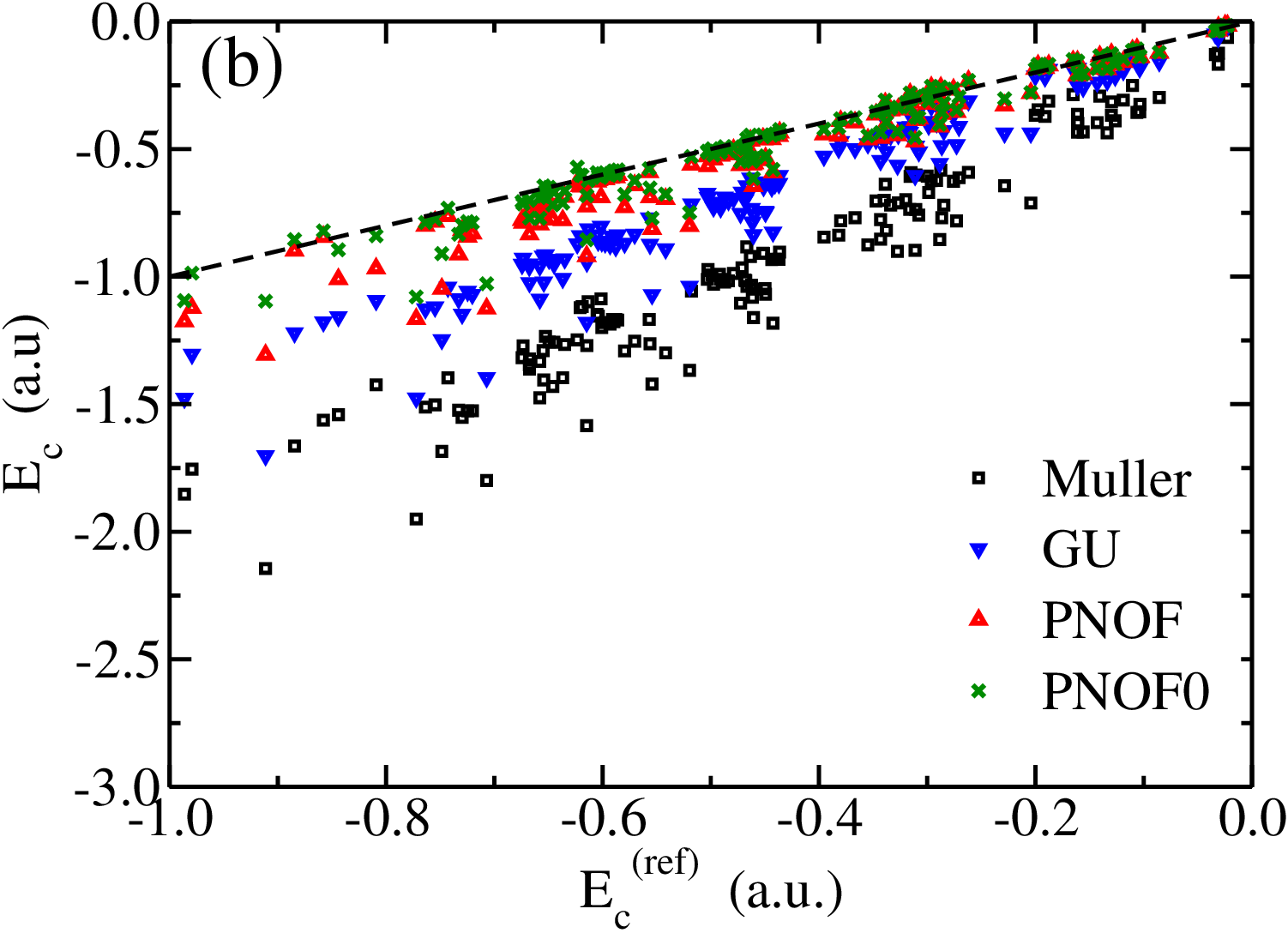} \\
          \\
          \\
        \includegraphics[width=7.5cm]{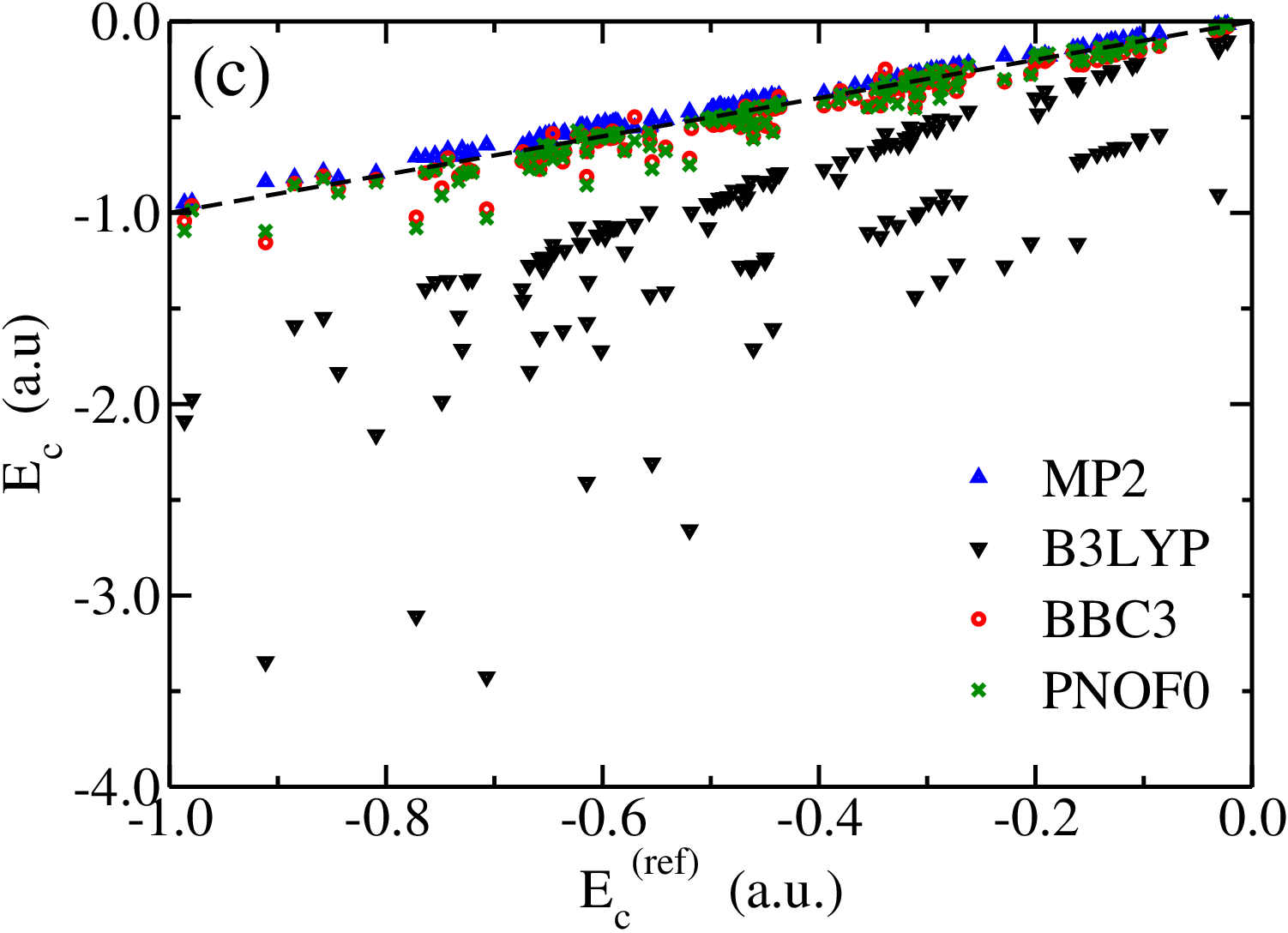} \\
     \end{tabular}
  \caption{ \label{fig:ec} 
    Correlation energy calculated by a variety of methods
    versus reference values ($E_{\rm c}^{\rm (ref)}$) obtained with CCSD(T). The
    dashed line corresponds to $E_c=E_c^{\rm (ref)}$. }
\end{figure}

In the Fig.~\ref{fig:ec}, we plot the correlation energy obtained with
various methods versus the reference values. The dashed lines
correspond to the perfect agreement with the reference energies
calculated with the CCSD(T) method. The complete set of our results
can be found as supplementary material to this Article.  It can be
easily seen from the figure that the M\"uller functional
overcorrelates substantially all systems. This is in complete
agreement with previous calculations for finite systems and for the HEG
and supports the idea that this functional could be a lower bound for
the total energy.~\cite{frank} In Figs.~\ref{fig:ec}(a) and (b), we
show the improvement over the M\"uller and GU functionals which is
provided by the BBC's and Piris functionals respectively.  In
Fig.~\ref{fig:ec}(c), we compare the results obtained with the more
accurate functionals, i.e. PNOF0 and BBC3, with MP2 and B3LYP values.

\begin{table*}[t!]
  \setlength{\tabcolsep}{0.5truecm}
  \centering
  \begin{tabular}{cllrll} 
    Method
    & \multicolumn{1}{c}{$\bar \Delta$} 
    & \multicolumn{1}{c}{$\Delta_{\rm max}$}
    & \multicolumn{1}{c}{$\bar \delta$} 
    & \multicolumn{1}{c}{$\delta_{\rm max}$}
    & \multicolumn{1}{c}{$\bar \delta_{\rm e}$}
    \\
    \hline \\[-2mm]
    M\"uller    & 0.55 & 1.23 (C$_2$Cl$_4$)  & 135.7\% & \ \ 438.3\% (Na$_2$) & 0.0193 \\
    GU          & 0.26 & 0.79 (C$_2$Cl$_4$)  & 51.63\% & \ \ 114.2\% (Si$_2$) & 0.0072 \\
    CGA         & 0.22 & 0.55 (C$_2$Cl$_4$)  & 69.11\% & \ \ 331.9\% (Na$_2$) & 0.0077 \\
    BBC1        & 0.29 & 0.75 (C$_2$Cl$_4$)  & 69.91\% & \ \ 159.1\% (Na$_2$) & 0.0098 \\ 
    BBC2        & 0.18 & 0.50 (C$_2$Cl$_4$)  & 45.02\% & \ \ 125.0\% (Na$_2$) & 0.0058 \\
    BBC3        & 0.068 & 0.27 (SiCl$_4$)    & 18.37\% &\ \ \ 50.8\% (SiH$_2$)& 0.0017 \\
    PNOF        & 0.102 & 0.42 (SiCl$_4$)    & 20.84\% &\ \ \ 59.1\% (SiCl$_4$)& 0.0021 \\
    PNOF0        & 0.072 & 0.32 (SiCl$_4$)    & 17.11\% &\ \ \ 46.0\% (Cl$_2$)  & 0.0015 \\
    \hline \\[-2mm]
    MP2         & 0.040 & 0.074 (C$_2$Cl$_4$)&  11.86\% &  \ \ \  35.7\% (Li$_2$) & 0.0015 \\
    B3LYP       & 0.75 &     2.72 (SiCl$_4$) & 305.0\% & 2803.7\% (Li$_2$) & 0.022 \\
    \hline
  \end{tabular}
  \caption{\label{tab:ec}
    Error in the correlation energies calculated with a variety of methods for the first set
    of calculations (whole G2/97 set, 6-31G* basis).
    The reference energies ($E_{\rm c}^{\rm ref}$) were obtained with CCSD(T). The values in the first, second and
    last column are in a.u.
  }
\end{table*}

\begin{table*}[t!]
  \setlength{\tabcolsep}{0.5truecm}
  \centering
  \begin{tabular}{cllrll} 
    Method
    & \multicolumn{1}{c}{$\bar \Delta$} 
    & \multicolumn{1}{c}{$\Delta_{\rm max}$}
    & \multicolumn{1}{c}{$\bar \delta$} 
    & \multicolumn{1}{c}{$\delta_{\rm max}$}
    & \multicolumn{1}{c}{$\bar \delta_{\rm e}$}
    \\
    \hline \\[-2mm]
    M\"uller    & 0.34  & 0.56 (Cl$_2$)  & 154.7\% & \ \ 438.8\% (Na$_2$) & 0.0191 \\
    GU          & 0.12  & 0.28 (Cl$_2$)  & 45.59\% & \ \ 120.4\% (Na$_2$) & 0.0049 \\
    CGA         & 0.16  & 0.33 (Si$_2$)  & 89.23\% & \ \ 330.9\% (Na$_2$) & 0.0085 \\
    BBC1        & 0.18  & 0.36 (ClO)     & 75.48\% & \ \ 180.8\% (Na$_2$) & 0.0096 \\ 
    BBC2        & 0.10  & 0.23 (Cl$_2$)  & 49.10\% & \ \ 144.6\% (Na$_2$) & 0.0055 \\
    BBC3        & 0.043 & 0.14 (Cl$_2$) & 19.98\% &\ \ \ 68.9\% (Na$_2$)& 0.0018 \\
    PNOF        & 0.046 & 0.16 (Cl$_2$) & 18.36\% &\ \ \ 49.5\% (Cl$_2$)& 0.0017 \\
    PNOF0       & 0.040 & 0.14 (Cl$_2$) & 16.84\% &\ \ \ 44.6\% (Na$_2$)  & 0.0016 \\
    \hline \\[-2mm]
    MP2         & 0.026 & 0.057 (Si$_2$H$_6$)&  14.55\% &  \ \ \  34.1\% (Li$_2$) & 0.0016 \\
    B3LYP       & 0.51  & 1.117 (Cl$_2$) & 423.5\% & 2642.5\% (Na$_2$) & 0.023 \\
    \hline
  \end{tabular}
  \caption{\label{tab:ec2}
    Error in the correlation energies calculated with a variety of methods for the second set
    of calculations (G2-1 set, cc-pVDZ basis). 
    The reference energies ($E_{\rm c}^{\rm ref}$) were obtained with CCSD(T). The values in the first, second and
    last column are in a.u.
  }
\end{table*}

In Tables~\ref{tab:ec} and \ref{tab:ec2}, we show a few quantities
that measure the deviation from the reference values.  The meaning of
the columns is the following:
\begin{subequations}
\begin{align}
  \bar \Delta &= \sqrt{\sum (E_{\rm c} - E_{\rm c}^{\rm (ref)})^2/N_{\rm mol}}\,, 
  \\ 
  \Delta_{\rm max} &=\max\left|E_{\rm c} - E_{\rm c}^{\rm (ref)}\right|\,,
  \\
  \bar \delta &= 100 \sqrt{\sum \left[(E_{\rm c} - E_{\rm c}^{\rm (ref)})/E_{\rm c}^{\rm (ref)}\right]^2/N_{\rm mol}}\,, \label{eq:delta}
  \\
  \delta_{\rm max}&=100\times\max\left|(E_{\rm c} - E_{\rm c}^{\rm (ref)})/E_{\rm c}^{\rm (ref)}\right|\,,
  \\
  \bar \delta_{\rm e} &=  \sum |E_{\rm c} - E_{\rm c}^{\rm (ref)}|/{(N\times N_{\rm mol})}\,,
\end{align}
\end{subequations}
where $N_{\rm mol}$ is the number of molecules in the test set, and
$N$ is the number of electrons in the molecule.

Clearly, the least performing approximation is the M\"uller
functional. On average it overestimates the correlation energy by
more than 100\% while in the worst case the error exceeds 400\%.  The
GU, CGA, BBC1 and BBC2 functionals improve significantly with errors
of the order of 45-90\% for both sets of calculations. Finally, BBC3
and the Piris functionals are the most accurate. Their performance is
quite remarkable, with average errors in the range 17-20\% which are
by merely a factor of 1.5 larger than the average error of MP2.  They
correct 85\% of the error of the M\"uller functional. Interestingly,
PNOF0, i.e. the Piris functional without the additional term which
keeps the occupations fractional, is slightly better in performance
than BBC3 or PNOF and is the most accurate as far as correlation
energies are concerned.

As we see in Fig.~\ref{fig:ec}(a), all 1-RDM functionals we tested
tend to overcorrelate compared to CCSD(T). This trend is the same for
both sets of calculations. For the first set, BBC3, PNOF and PNOF0
give total energies lower than CCSD(T) for 120, 126 and 104 systems
respectively out of the 148 of the G2 set. The other 1-RDM functionals
overcorrelate for nearly all the cases.

Note that the performance of the functionals for the two sets of
calculations is quite similar (see Tables~\ref{tab:ec}
and~\ref{tab:ec2}). This indicates that the convergence rate of the
correlation energies obtained with 1-RDM functionals is similar to
variational methods like MP2 or CCSD(T). Note that it is well known
that DFT functionals converge faster with respect to the basis set
faster than these variational methods.~\cite{gil,martin}

In the last column of Tables~\ref{tab:ec} and~\ref{tab:ec2}, we show
the average error, $\bar{\delta}_e$, of the correlation energy per electron. 
The largest
value is for the M\"uller functional, 0.019 a.u., and is one order of
magnitude smaller than the correction of 1/8 a.u. per electron
proposed by Frank {\it et al}.~\cite{frank} Thus, it is clear that
this correction is rather unrealistic leading to very high total
energies.

We now turn our attention to the quality of the minimizing
$\gamma$. We have already mentioned that most of the functionals
considered in this work (all except PNOF) produce pinned states for
all systems except those with only two electrons. The presence of a
pinned state means that the particular natural orbital is present in
all determinants of the full configuration-interaction expansion with
non-zero coefficients. This situation is rather unlikely for the
systems included in the present benchmark.  Nevertheless, we would not
consider it necessarily as a serious drawback.  Pinned states are
usually core states with occupation extremely close to one also in the
exact $\rdm$. Moreover, these core states do not affect significantly
many important quantities based on total energy differences.

\begin{table*}
\setlength{\tabcolsep}{0.2truecm}
\centering
\begin{tabular}{lccccccccc}
Molecule &	M\"uller& GU &   CGA &   BBC1&	BBC2&	BBC3&	PNOF & PNOF0 & CCD    \\
\hline\\[-2mm]
Ammonia	 &	0.246&	0.080&	0.219&	0.153&	0.126&	0.071& 0.053 & 0.052 &	0.061\\
Ethane	 &	0.379&	0.183&	0.309&	0.126&	0.154&	0.093& 0.106 & 0.095 &	0.112\\
CO$_2$	 &	0.560&	0.276&	0.520&	0.329&	0.229&	0.149& 0.167 & 0.154 &	0.131\\
H$_2$O	 &	0.220&	0.066&	0.200&	0.148&	0.114&	0.065& 0.047 & 0.048 &	0.055\\
Acetylene &	0.447&	0.183&	0.435&	0.239&	0.191&	0.108& 0.088 & 0.087 &	0.115\\
\hline\\[-2mm]
$\bar{\delta}$ (\%) &  293& 	65&	258&	132&	82&	15 & 18 & 17  \\
\hline
\end{tabular}
\caption{\label{tab:correl_q} 
Total charge $q_w$ (electrons per spin) occupying weakly occupied
natural orbitals for a few representative molecules, compared with the
result of coupled cluster doubles (CCD).}
\end{table*}

A quantity which is a measure, although still not absolute, of the
quality of the occupation numbers produced by the 1-RDM functionals is
the total charge $q_w$ occupying the weakly occupied states of the set
$W$. The quantity $q_w$ shows how much the calculated $\rdm$ differs
from an idempotent matrix, i.e., it is a measure of the degree of
correlation of the system. We calculated this quantity for a few
representative molecules using the 1-RDM functionals. The results are
in the Table~\ref{tab:correl_q} where we have also included the values
obtained with coupled cluster with double excitations
(CCD)~\cite{ccd}.  For the CCD calculations, we used the Gaussian 98
program employing the same basis set as above.  The overcorrelation of
the M\"uller functional is again obvious, overestimating $q_w$ by a
factor of 4.  The rest of the functionals improve significantly as far
as this quantity is concerned, with BBC3 being the most accurate, with
an error of only 15\%. The second best is the PNOF0 functional with an
error of 17\%.

We would like to close the discussion on the occupation numbers by
mentioning a minor problem of the BBC3 functional. This functional
introduces orbital specific corrections involving the bonding and
anti-bonding orbitals. These corrections are repulsive, i.e., they
reduce (in absolute value) the correlation energy. As a consequence,
the tendency in the minimization process is to lower the occupation of
these states and in that way counteract the effect of the repulsive
correction.  In several cases, the impact of this counteraction is so
strong that the occupation number of some of the anti-bonding orbitals
gets equal to zero (pinned at zero). One remedy that we tried for this
problem was to update, during the variation, the grouping of the
orbitals into the 4 sets mentioned in the previous section.  This
involves a mechanism that repeatedly finds the anti-bonding orbitals,
for example, according to their occupation. However, this process adds
a non-analytic behavior to the total energy as a function of the
occupation numbers. For this reason, it was not possible to converge
the calculation for several molecules, which lead us to abandon this
path. Thus, the version of BBC3 functional that we used for all our
calculations is the one of Eq.~(\ref{eq:BBC3}), with the grouping of
the orbitals into the 4 different sets defined at the beginning from
the HF initial guess.  However, it is very common that some of the
orbitals chosen as anti-bonding are left with zero or almost zero
occupation for the optimal $\rdm$.

%%%%%%%%%%%%%%%%%%%%%%%%%%%%%%%%%%%%%%%%%%%%%%%%%%%%%%%%%%%%%%%%%%%%%%%%%%%
\section{Atomization energies}

\begin{table*}
\setlength{\tabcolsep}{0.2truecm}
\centering
\begin{tabular}{cccccccccccc}
\hline
Calculation &     R(O)HF & MP2  & B3LYP & Mueller &  GU  & CGA  & BBC1 & BBC2  & BBC3 & PNOF0 & PNOF \\
\hline
set 1       &     42.4   & 6.24 &  11.7 &  32.7   & 43.7 & 42.5 & 31.0 & 26.9  & 18.0 & 17.5  & 25.5 \\
set 2       &     53.8   & 7.94 &  12.1 &  40.6   & 50.4 & 47.4 & 34.8 & 40.1  & 25.6 & 23.9  & 30.4 \\      
\hline
\end{tabular}
\caption{\label{tab:atomization}
The error average error $\bar \delta$ (\%), defined in Eq.~\ref{eq:delta}, for the atomization energies calculated with different methods for
the two sets of calculations. Reference values are the atomization energies calculated with CCSD(T) method.
}
\end{table*}

It is clear from Tables~\ref{tab:ec}, and \ref{tab:ec2} that all 1-RDM
functionals give a better account for the correlation energy than
B3LYP (which strongly overcorrelates).  However, for methods that are
not variational, like RDMFT or DFT, one should not fall in the
temptation of using the quality of the calculated correlation energy
as the only criterion to assess the performance of functionals. In
fact, quantities that are total energy differences are sometimes much
more important than the correlation energy. For this reason we
also included atomization energies in our benchmark.

For this task, we calculated the energies of the atoms with the same
functionals and basis sets used for the molecules (these results can
also be found as supplementary material). The large majority of the
atoms involved are open shell systems. Like in the case of the
open shell systems of the G2/97 set, we employed the open shell
generalization introduced in
Ref.~\onlinecite{our_open_shell}. Clearly, the performance of the
functionals in the calculation of the atomization energies is also
affected by this choice.

Tables with atomization energies of all functionals are included as
supplementary material to this Article.  The performance of all the
methods is summarized in Table~\ref{tab:atomization}. As in the case
of total energies, the latest generation functionals, i.e. BBC3 and
Piris functionals, perform better compared to their
predecessors. Again, in this case, PNOF0 gives slightly better results
than BBC3 and PNOF. Even if the performance of the Piris and BBC3
functionals is quite satisfactory, they still do not provide the
accuracy of MP2 or even B3LYP. The error of PNOF0 is 1.5 times the
error of B3LYP and about 3 times the error of MP2. Furthermore, the
BBC3 functional yields a negative atomization energy (unbound system)
for Li$_2$. This is probably due to a size-inconsistency problem of
the open shell treatment. 
Indeed, most of the atoms involved are open shell systems. 
It is a question whether the sum of the total energies of the atoms 
consisting a molecule equals the total energy of the stretched molecule 
obtained by a single closed-shell calculation.
Li$_2$ is an extreme case, which is also
enhanced by the very small value of its atomization energy.
However, more atomization energies might deviate from the
exact due to this inconsistency. It would be interesting to
investigate if different open shell schemes, like for example that of
Ref.~\onlinecite{piris_os}, improve the results of RDMFT on the
atomization energies.

\section{Conclusions}
\label{sec:concl}

We presented benchmark calculations on an extended set of molecules
for a series of exchange-correlation functionals proposed within
RDMFT. We assessed the accuracy of these functionals
concerning both the correlation and the atomization energy.

All functionals fared quite well in the test, with errors in the
correlation energy at least an order of magnitude smaller than B3LYP,
perhaps the most used functional in the world of DFT. To summarize,
the oldest of the approximations within RDMFT, the functional of
M\"uller, overcorrelates substantially. The GU and CGA improve
significantly compared to the functional of M\"uller.  Finally, the
BBC3 functional of Gritsenko {\it et al.}, and the functional of Piris
are the most successful in correcting the error of the M\"uller
functional. PNOF0, i.e. the Piris functional {\it without} the term
which prevents pinned states, performs better than all the functional
we considered, with a precision already comparable to MP2 theory.

For the atomization energies, 1-RDM functionals are also satisfactory,
yielding however a somewhat lower accuracy than B3LYP. The best 1-RDM
functional for this quantity is again PNOF0, with an average error 1.5
times larger than B3LYP. We found that the treatment of open shell
atoms introduces a size inconsistency which affects significantly the
accuracy of the atomization energies.

We also report a problem in the behavior of BBC3 functional which
comes from the orbital-specific corrections introduced in this
functional. More specifically, we observed a counteraction to the
repulsive correction, that leads to zero occupation for some of the
anti-bonding orbitals. In spite of this problem, BBC3, as well as the
Piris functional, was found very accurate in reproducing the total
charge that occupies the weakly occupied states.

It is clear from our results that the available functionals already
have enough precision to be used in many quantum chemistry problems,
especially for systems that DFT results are not satisfactory.
Furthermore, RDMFT is still a young field, and it is reasonable to
expect fast advances in the construction of new functionals, with
increased precision and range of applicability. However, one big
problem still remains to be solved: the efficiency of the minimization
of the total energy with respect to the 1-RDM. In addition, up to now,
the existing RDMFT codes are research codes, relatively slow, and of
limited availability. We hope that the results present in this paper
will motivate developers of mainstream quantum chemistry codes to
include RDMFT in their set of methods, bringing therefore this very
promising approach to a much wider public.

\begin{acknowledgments}
  This work was supported in part by the Deutsche Forschungsgemeischaft within
  the program SPP 1145, by the Portuguese FCT through the project
  PTDC/FIS/73578/2006, and by the EC Network of Excellence NANOQUANTA
  (NMP4-CT-2004-500198).  Part of the calculations were performed at the
  Laborat\'orio de Computa\c{c}\~ao Avan\c{c}ada of the University of Coimbra.
\end{acknowledgments}

%\bibliography{hp}

\begin{widetext}

\newpage
\centerline{\LARGE\bf Suplementary material}

\setcounter{section}{0}
\setcounter{table}{0}
\section{Correlation energies}
\subsection{G2/97 set, 6-31G* basis}

\begin{longtable*}{p{5.4cm}dddddd}
\setlength{\tabcolsep}{0.5truecm}
{\bf System} & 
\multicolumn{1}{c}{$-E_{\rm tot}$ RHF} & 
\multicolumn{1}{c}{$-E_{\rm c}$ CCSD(T)} & 
\multicolumn{1}{c}{$-E_{\rm c}$ BBC1} & 
\multicolumn{1}{c}{$-E_{\rm c}$ BBC3} & 
\multicolumn{1}{c}{$-E_{\rm c}$ PNOF0}& 
\multicolumn{1}{c}{$-E_{\rm c}$ PNOF} \\ 
\hline \\[-2mm]
\endhead
LiH						&    7.98087 &  0.0226 &  0.0393 & 0.0266 & 0.0161 & 0.0161 \\
BeH						&   15.14697 &  0.0339 &  0.0764 & 0.0473 & 0.0380 & 0.0380 \\
CH radical					&   38.26128 &  0.1063 &  0.2507 & 0.1210 & 0.1074 & 0.1083 \\
CH$_2$ ($^3$B$_1$)				&   38.91631 &  0.1104 &  0.1806 & 0.1226 & 0.1129 & 0.1134 \\
CH$_2$ ($^1$A$_1$)				&   38.87219 &  0.1299 &  0.2651 & 0.1489 & 0.1266 & 0.1275 \\
CH$_3$ radical					&   39.55461 &  0.1406 &  0.2127 & 0.1468 & 0.1344 & 0.1363 \\
CH$_4$						&   40.19507 &  0.1653 &  0.2187 & 0.1637 & 0.1490 & 0.1555 \\
NH						&   54.95210 &  0.1297 &  0.2356 & 0.1410 & 0.1297 & 0.1316 \\
NH$_2$ radical					&   55.55299 &  0.1613 &  0.2766 & 0.1711 & 0.1549 & 0.1567 \\
NH$_3$						&   56.18384 &  0.1912 &  0.2908 & 0.2077 & 0.1811 & 0.1819 \\
OH radical					&   75.37792 &  0.1612 &  0.3171 & 0.1779 & 0.1592 & 0.1606 \\
H$_2$O						&   76.00981 &  0.2002 &  0.3017 & 0.2131 & 0.1861 & 0.1857 \\
HF						&  100.00229 &  0.1878 &  0.2647 & 0.1921 & 0.1685 & 0.1716 \\
SiH$_2$ ($^1$A$_1$)				&  289.99970 &  0.1038 &  0.2438 & 0.1546 & 0.1382 & 0.1401 \\
SiH$_2$ ($^3$B$_1$)				&  289.99090 &  0.0856 &  0.1995 & 0.1291 & 0.1210 & 0.1225 \\
SiH$_3$ radical					&  290.60476 &  0.1034 &  0.2185 & 0.1465 & 0.1348 & 0.1375 \\
SiH$_4$						&  291.22504 &  0.1189 &  0.2237 & 0.1611 & 0.1469 & 0.1586 \\
PH$_2$ radical					&  341.84526 &  0.1263 &  0.2753 & 0.1751 & 0.1668 & 0.1683 \\
PH$_3$						&  342.44775 &  0.1430 &  0.2870 & 0.2036 & 0.1848 & 0.1861 \\
H$_2$S						&  398.66711 &  0.1559 &  0.3249 & 0.2270 & 0.2057 & 0.2068 \\
HCl						&  460.05985 &  0.1608 &  0.3145 & 0.2262 & 0.2082 & 0.2112 \\
Li$_2$						&   14.86689 &  0.0310 &  0.0593 & 0.0340 & 0.0230 & 0.0230 \\
LiF						&  106.93418 &  0.1976 &  0.2777 & 0.1902 & 0.1681 & 0.1726 \\
C$_2$H$_2$					&   76.81560 &  0.2874 &  0.3946 & 0.2625 & 0.2569 & 0.2589 \\
C$_2$H$_4$					&   78.03107 &  0.3001 &  0.4317 & 0.2860 & 0.2869 & 0.2900 \\
C$_2$H$_6$					&   79.22854 &  0.3150 &  0.4314 & 0.3118 & 0.3038 & 0.3182 \\
CN radical					&   92.19013 &  0.2963 &  0.4362 & 0.2738 & 0.2542 & 0.2582 \\
HCN						&   92.87019 &  0.3159 &  0.4313 & 0.2783 & 0.2805 & 0.2839 \\
CO						&  112.73448 &  0.3104 &  0.4524 & 0.3226 & 0.2931 & 0.3023 \\
HCO  radical					&  113.24033 &  0.3199 &  0.5100 & 0.2852 & 0.3175 & 0.3230 \\
H$_2$CO						&  113.86371 &  0.3339 &  0.5375 & 0.3640 & 0.3411 & 0.3489 \\
H$_3$COH					&  115.03419 &  0.3474 &  0.5379 & 0.3782 & 0.3542 & 0.3644 \\
N$_2$						&  108.93540 &  0.3388 &  0.4781 & 0.2493 & 0.3086 & 0.3206 \\
H$_2$NNH$_2$					&  111.16800 &  0.3668 &  0.5762 & 0.4027 & 0.3770 & 0.3957 \\
NO radical					&  129.24050 &  0.3431 &  0.5929 & 0.2999 & 0.3513 & 0.3629 \\
O$_2$						&  149.58560 &  0.3800 &  0.6101 & 0.3619 & 0.3814 & 0.3901 \\
HOOH						&  150.76011 &  0.3955 &  0.6729 & 0.4397 & 0.4217 & 0.4426 \\
F$_2$						&  198.67283 &  0.3818 &  0.6925 & 0.4304 & 0.4150 & 0.4471 \\
CO$_2$						&  187.62841 &  0.5032 &  0.7321 & 0.5140 & 0.5084 & 0.5290 \\
Na$_2$						&  323.68169 &  0.0312 &  0.0808 & 0.0438 & 0.0361 & 0.0362 \\
Si$_2$						&  577.69210 &  0.2044 &  0.4507 & 0.2719 & 0.2786 & 0.2808 \\
P$_2$						&  681.42041 &  0.2729 &  0.5286 & 0.3648 & 0.3464 & 0.3549 \\
S$_2$						&  795.00558 &  0.2885 &  0.6032 & 0.3636 & 0.4031 & 0.4115 \\
Cl$_2$						&  918.91256 &  0.3112 &  0.6721 & 0.4446 & 0.4544 & 0.4704 \\
NaCl						&  621.39961 &  0.1612 &  0.3246 & 0.2227 & 0.2054 & 0.2097 \\
SiO						&  363.77506 &  0.2987 &  0.4877 & 0.3175 & 0.3116 & 0.3226 \\
CS						&  435.30360 &  0.2847 &  0.5075 & 0.3341 & 0.3201 & 0.3260 \\
SO						&  472.30865 &  0.3378 &  0.6018 & 0.3635 & 0.3946 & 0.3994 \\
ClO radical					&  534.22659 &  0.3275 &  0.6872 & 0.3872 & 0.4311 & 0.4455 \\
ClF						&  558.81813 &  0.3432 &  0.6596 & 0.4408 & 0.4355 & 0.4560 \\
Si$_2$H$_6$					&  581.30488 &  0.2286 &  0.4445 & 0.3151 & 0.3025 & 0.3289 \\
CH$_3$Cl					&  499.09290 &  0.3106 &  0.5496 & 0.3898 & 0.3753 & 0.3862 \\
H$_3$C-SH					&  437.70000 &  0.3077 &  0.5567 & 0.3925 & 0.3762 & 0.3805 \\
HOCl						&  534.84018 &  0.3549 &  0.6763 & 0.4458 & 0.4477 & 0.4620 \\
SO$_2$						&  547.15790 &  0.5562 &  0.9288 & 0.6249 & 0.6521 & 0.6890 \\
BF$_3$						&  323.19407 &  0.6133 &  0.8385 & 0.6169 & 0.6238 & 0.6779 \\
BCl$_3$						& 1403.25416 &  0.5543 &  1.0356 & 0.7339 & 0.7708 & 0.8128 \\
AlF$_3$						&  540.44920 &  0.6014 &  0.8175 & 0.6039 & 0.6109 & 0.6886 \\
AlCl$_3$ 					& 1620.57601 &  0.5195 &  0.9946 & 0.7154 & 0.7490 & 0.8018 \\
CF$_4$						&  435.64152 &  0.8440 &  1.2196 & 0.8742 & 0.8950 & 1.0098 \\
CCl$_4$						& 1875.74481 &  0.7721 &  1.4797 & 1.0231 & 1.0794 & 1.1659 \\
O=C=S						&  510.25587 &  0.4725 &  0.7867 & 0.5531 & 0.5489 & 0.5627 \\
CS$_2$						&  832.88348 &  0.4425 &  0.8240 & 0.5687 & 0.5798 & 0.5917 \\
COF$_2$						&  311.61034 &  0.6744 &  1.0000 & 0.7292 & 0.7064 & 0.7777 \\
SiF$_4$						&  686.94780 &  0.8092 &  1.1123 & 0.8257 & 0.8401 & 0.9679 \\
SiCl$_4$					& 2127.04685 &  0.7071 &  1.3695 & 0.9798 & 1.0279 & 1.1252 \\
N$_2$O  					&  183.66310 &  0.5569 &  0.8539 & 0.5872 & 0.5772 & 0.5907 \\
ClNO						&  588.67097 &  0.5418 &  0.9989 & 0.6571 & 0.6766 & 0.6953 \\
NF$_3$						&  352.53192 &  0.7329 &  1.2001 & 0.8089 & 0.8346 & 0.9132 \\
PF$_3$						&  639.12725 &  0.6675 &  1.0267 & 0.7519 & 0.7705 & 0.8337 \\
O$_3$						&  224.23806 &  0.6459 &  1.1124 & 0.5865 & 0.7192 & 0.7762 \\
F$_2$O						&  273.44465 &  0.5796 &  1.0466 & 0.6707 & 0.6779 & 0.7274 \\
ClF$_3$						&  757.47982 &  0.7485 &  1.2977 & 0.8695 & 0.9099 & 1.0476 \\
C$_2$F$_4$					&  473.41166 &  0.9863 &  1.4781 & 1.0433 & 1.0943 & 1.1753 \\
C$_2$Cl$_4$					& 1913.60266 &  0.9114 &  1.6591 & 1.1556 & 1.0961 & 1.3067 \\
CF$_3$CN  					&  428.47216 &  0.9793 &  1.3702 & 0.9630 & 0.9863 & 1.1229 \\
CH$_3$CCH (propyne)				&  115.86191 &  0.4364 &  0.6169 & 0.4477 & 0.4233 & 0.4349 \\
CH$_2$=C=CH$_2$ (allene)			&  115.86012 &  0.4371 &  0.6340 & 0.3925 & 0.4360 & 0.4482 \\
C$_3$H$_4$ (cyclopropene)			&  115.82178 &  0.4407 &  0.6384 & 0.4568 & 0.4402 & 0.4466 \\
CH$_3$CH=CH$_2$ (propylene)			&  117.07068 &  0.4512 &  0.6511 & 0.4439 & 0.4523 & 0.4618 \\
C$_3$H$_6$ (cyclopropane)			&  117.05853 &  0.4518 &  0.6440 & 0.4637 & 0.4445 & 0.4598 \\
C$_3$H$_8$ (propane)				&  118.26328 &  0.4668 &  0.6428 & 0.4581 & 0.4589 & 0.4862 \\
CH$_2$CHCHCH$_2$ (Trans-1,3-butadiene)		&  154.91815 &  0.5878 &  0.8490 & 0.6006 & 0.5986 & 0.6106 \\
C$_4$H$_6$ (2-butyne)				&  154.90663 &  0.5858 &  0.8340 & 0.5997 & 0.5818 & 0.6018 \\
C$_4$H$_6$ (methylene cyclopropane)		&  154.88650 &  0.5895 &  0.8370 & 0.6077 & 0.5855 & 0.6081 \\
C$_4$H$_6$ (bicyclobutane)		 	&  154.87081 &  0.5931 &  0.8525 & 0.6114 & 0.6030 & 0.6127 \\
C$_4$H$_6$ (cyclobutene)			&  154.89853 &  0.5907 &  0.8365 & 0.5725 & 0.5883 & 0.6023 \\
C$_4$H$_8$ (cyclobutane)			&  156.09645 &  0.6050 &  0.8414 & 0.5884 & 0.5923 & 0.6272 \\
C$_4$H$_8$ (isobutene)				&  156.10979 &  0.6041 &  0.8554 & 0.6245 & 0.6047 & 0.6219 \\
C$_4$H$_{10}$ (trans-butane)			&  157.29791 &  0.6188 &  0.8422 & 0.5943 & 0.6037 & 0.6468 \\
C$_4$H$_{10}$ (isobutane)			&  157.29841 &  0.6207 &  0.8438 & 0.5941 & 0.6020 & 0.6448 \\
C$_5$H$_8$ (spiropentane)			&  193.91719 &  0.7428 &  1.0339 & 0.7131 & 0.7314 & 0.7629 \\
C$_6$H$_6$ (benzene)				&  230.70204 &  0.8579 &  1.1510 & 0.8076 & 0.8205 & 0.8440 \\
H$_2$CF$_2$					&  237.89483 &  0.5026 &  0.7506 & 0.5123 & 0.5139 & 0.5675 \\
HCF$_3$						&  336.76920 &  0.6734 &  0.9854 & 0.6811 & 0.7150 & 0.7888 \\
H$_2$CCl$_2$					&  957.98494 &  0.4606 &  0.8580 & 0.6005 & 0.6145 & 0.6454 \\
HCCl$_3$					& 1416.86955 &  0.6147 &  1.1725 & 0.8103 & 0.8556 & 0.9198 \\
H$_3$C-NH$_2$ (methylamine)			&   95.20912 &  0.3398 &  0.5236 & 0.3744 & 0.3463 & 0.3551 \\
CH$_3$-CN (acetonitrile)			&  131.92248 &  0.4634 &  0.6507 & 0.4690 & 0.4455 & 0.4666 \\
CH$_3$-NO$_2$ (nitromethane)			&  243.65373 &  0.7246 &  1.1502 & 0.7714 & 0.7844 & 0.8438 \\
CH$_3$ONO (methylnitrite)			&  243.65962 &  0.7203 &  1.1650 & 0.7821 & 0.7891 & 0.8306 \\
CH$_3$-SiH$_3$ (methylsilane)			&  330.27223 &  0.2707 &  0.4371 & 0.3134 & 0.2956 & 0.3206 \\
HCOOH (formic acid)				&  188.75863 &  0.5178 &  0.7936 & 0.5591 & 0.5292 & 0.5608 \\
HCOOCH$_3$ (methyl formate)			&  227.78570 &  0.6677 &  1.0141 & 0.7131 & 0.7012 & 0.7475 \\
CH$3$CONH$_2$ (acetamide)			&  207.97351 &  0.6580 &  1.0018 & 0.7057 & 0.6905 & 0.7270 \\
C$_2$H$_4$NH (aziridine)			&  133.03730 &  0.4790 &  0.7179 & 0.5155 & 0.4915 & 0.5035 \\
NCCN (cyanogen)   				&  184.57786 &  0.6233 &  0.8564 & 0.5981 & 0.5710 & 0.6297 \\
(CH$_3$)$_2$NH (dimethylamine) 			&  134.23800 &  0.4909 &  0.7533 & 0.5396 & 0.5136 & 0.5261 \\
CH$_3$CH$_2$NH$_2$ (trans-ethylamine)		&  134.24677 &  0.4920 &  0.7403 & 0.5339 & 0.5068 & 0.5223 \\
H$_2$CCO (ketene)				&  151.72173 &  0.4685 &  0.7296 & 0.5144 & 0.4995 & 0.5081 \\
C$_2$H$_4$O (oxirane)				&  152.86540 &  0.4873 &  0.7475 & 0.5301 & 0.4905 & 0.5182 \\
CH$_3$CHO (acetaldehyde)			&  152.91350 &  0.4837 &  0.7557 & 0.5259 & 0.5050 & 0.5212 \\
OCHCHO (glyoxal)				&  226.58645 &  0.6543 &  1.0334 & 0.6923 & 0.6448 & 0.7629 \\
CH$_3$CH$_2$OH (ethanol)			&  154.07437 &  0.4989 &  0.7473 & 0.5349 & 0.4988 & 0.5397 \\
CH$_3$-O-CH$_3$	(dimethylether)			&  154.06337 &  0.4975 &  0.7694 & 0.5424 & 0.5233 & 0.5399 \\
C$_2$H$_4$S (thiooxirane)			&  475.54647 &  0.4494 &  0.7605 & 0.5466 & 0.5268 & 0.5354 \\
(CH$_3$)$_2$SO (dimethylsulfoxide)		&  551.53596 &  0.6580 &  1.0980 & 0.7725 & 0.7719 & 0.7959 \\
C$_2$H$_5$SH (ethanethiol)			&  476.73523 &  0.4600 &  0.7652 & 0.5526 & 0.5368 & 0.5558 \\
CH$_3$-S-CH$_3$ (dimethyl sulfide)		&  476.73487 &  0.4615 &  0.7863 & 0.5614 & 0.5480 & 0.5591 \\
H$_2$C=CHF (vinylfluoride)			&  176.88072 &  0.4711 &  0.7073 & 0.5060 & 0.4918 & 0.5049 \\
C$_2$H$_5$Cl (ethylchloride)			&  538.13118 &  0.4624 &  0.7625 & 0.5425 & 0.5340 & 0.5604 \\
CH$_2$=CHCl (vinylchloride)			&  536.93293 &  0.4498 &  0.7563 & 0.5383 & 0.5316 & 0.5438 \\
H$_2$C=CHCN (acrylonitrile)			&  169.76199 &  0.6007 &  0.8470 & 0.5996 & 0.5962 & 0.6140 \\
CH$_3$-CO-CH$_3$ (acetone)			&  191.95987 &  0.6349 &  0.9647 & 0.6771 & 0.6635 & 0.6871 \\
CH$_3$COOH (acetic acid)			&  227.80703 &  0.6675 &  1.0068 & 0.7119 & 0.6984 & 0.7394 \\
CH$_3$COF (acetyl fluoride)			&  251.79498 &  0.6549 &  0.9765 & 0.6987 & 0.6848 & 0.7344 \\
CH$_3$COCl (acetyl chloride)			&  611.82872 &  0.6368 &  1.0434 & 0.7335 & 0.7117 & 0.7792 \\
CH$_3$CH$_2$CH$_2$Cl (propyl chloride)		&  577.16609 &  0.6145 &  0.9589 & 0.6794 & 0.6827 & 0.7245 \\
(CH$_3$)$_2$CHOH (isopropanol)			&  193.11389 &  0.6528 &  0.9504 & 0.6771 & 0.6643 & 0.7036 \\
C$_2$H$_5$-O-CH$_3$ (methyl-ethyl ether)	&  193.10336 &  0.6494 &  0.9651 & 0.6845 & 0.6500 & 0.7085 \\
(CH$_3$)$_3$N (trimethylamine)			&  173.26829 &  0.6448 &  0.9659 & 0.6905 & 0.6698 & 0.6879 \\
C$_4$H$_4$O (furan)				&  228.62241 &  0.7636 &  1.1039 & 0.7917 & 0.7879 & 0.8007 \\
C$_4$H$_4$S (thiophene)				&  551.28822 &  0.7297 &  1.1178 & 0.7936 & 0.7983 & 0.8178 \\
C$_4$H$_4$NH (pyrrole)				&  208.80589 &  0.7548 &  1.0905 & 0.7773 & 0.7743 & 0.7852 \\
C$_5$H$_5$N (pyridine)				&  246.69375 &  0.8846 &  1.2240 & 0.8461 & 0.8549 & 0.8980 \\
H$_2$						&    1.12679 &  0.0249 &  0.0262 & 0.0251 & 0.0150 & 0.0150 \\
SH radical					&  398.06041 &  0.1338 &  0.3247 & 0.1861 & 0.1858 & 0.1872 \\
CCH radical					&   76.13413 &  0.2619 &  0.3899 & 0.2574 & 0.2309 & 0.2329 \\
C$_2$H$_3$ radical ($^2$A$'$)			&   77.38072 &  0.2757 &  0.4276 & 0.2605 & 0.2649 & 0.2668 \\
CH$_3$CO radical ($^2$A$'$)			&  152.29117 &  0.4677 &  0.7365 & 0.4427 & 0.4802 & 0.5011 \\
H$_2$COH radical ($^2$A)			&  114.40397 &  0.3268 &  0.5256 & 0.3487 & 0.3355 & 0.3428 \\
CH$_3$O radical (Cs)				&  114.41185 &  0.3163 &  0.5554 & 0.3346 & 0.3348 & 0.3451 \\
CH$_3$CH$_2$O radical ($^2$A$''$)		&  153.45188 &  0.4665 &  0.7641 & 0.4872 & 0.4943 & 0.5181 \\
CH$_3$S radical ($^2$A$'$)			&  437.09727 &  0.2866 &  0.5559 & 0.3446 & 0.3563 & 0.3620 \\
C$_2$H$_5$ radical ($^2$A$'$)			&   78.59242 &  0.2913 &  0.4383 & 0.3026 & 0.2891 & 0.2951 \\
(CH$_3$)$_2$CH radical ($^2$A$'$)		&  117.63116 &  0.4437 &  0.6640 & 0.4575 & 0.4462 & 0.4616 \\
(CH$_3$)$_3$C (t-butyl radical)			&  156.66989 &  0.5975 &  0.8760 & 0.6006 & 0.5954 & 0.6181 \\
NO$_2$ radical					&  204.01248 &  0.5703 &  0.9422 & 0.4991 & 0.6231 & 0.6416 \\
\hline\\[4mm]
\caption{Correlation energies $E_c$ for the full G2/97 test set calculated with
the 6-31G* basis. 
All quantities are in a.u.}
\label{tab:correl} \\
\end{longtable*}

\subsection{G2-1 set, cc-pVDZ basis}

%\begin{longtable*}{p{5.2cm}dddddd}
\begin{longtable*}{p{5.4cm}dddddd}
\setlength{\tabcolsep}{0.5truecm}
{\bf System} & 
\multicolumn{1}{c}{$-E_{\rm tot}$ RHF} & 
\multicolumn{1}{c}{$-E_{\rm c}$ CCSD(T)} & 
\multicolumn{1}{c}{$-E_{\rm c}$ BBC1} & 
\multicolumn{1}{c}{$-E_{\rm c}$ BBC3} & 
\multicolumn{1}{c}{$-E_{\rm c}$ PNOF0}& 
\multicolumn{1}{c}{$-E_{\rm c}$ PNOF} \\ 
\hline \\[-2mm]
\endhead
LiH						& 7.98367    & 0.0324 & 0.0506 &0.0354 & 0.0241& 0.0241  \\
BeH						& 15.14952   & 0.0409 & 0.0790 &0.0524 & 0.0405& 0.0406  \\
CH radical					& 38.26903   & 0.1146 & 0.2511 &0.1207 & 0.1087& 0.1097  \\
CH$_2$ ($^3$B$_1$)				& 38.92170   & 0.1216 & 0.1898 &0.1303 & 0.1188& 0.1200  \\
CH$_2$ ($^1$A$_1$)				& 38.88109   & 0.1429 & 0.2709 &0.1534 & 0.1336& 0.1345  \\
CH$_3$ radical					& 39.55963   & 0.1585 & 0.2266 &0.1579 & 0.1448& 0.1469  \\
CH$_4$						& 40.19871   & 0.1912 & 0.2381 &0.1795 & 0.1639& 0.1710  \\
NH						& 54.95987   & 0.1363 & 0.2413 &0.1447 & 0.1343& 0.1363  \\
NH$_2$ radical					& 55.56301   & 0.1750 & 0.2863 &0.1791 & 0.1628& 0.1650  \\
NH$_3$						& 56.19561   & 0.2123 & 0.3050 &0.2205 & 0.1930& 0.1939  \\
OH radical					& 75.39003   & 0.1747 & 0.3270 &0.1882 & 0.1711& 0.1725  \\
H$_2$O						& 76.02638   & 0.2206 & 0.3140 &0.2267 & 0.2000& 0.2015  \\
HF						& 100.01888  & 0.2151 & 0.2867 &0.2160 & 0.1932& 0.1962  \\
SiH$_2$ ($^1$A$_1$)				& 290.01863  & 0.1282 & 0.2562 &0.1655 & 0.1492& 0.1511  \\
SiH$_2$ ($^3$B$_1$)				& 290.00997  & 0.1068 & 0.2107 &0.1393 & 0.1306& 0.1323  \\
SiH$_3$ radical					& 290.62377  & 0.1342 & 0.2370 &0.1622 & 0.1495& 0.1527  \\
SiH$_4$						& 291.24305  & 0.1600 & 0.2461 &0.1843 & 0.1681& 0.1809  \\
PH$_2$ radical					& 341.86779  & 0.1513 & 0.2891 &0.1835 & 0.1793& 0.1806  \\
PH$_3$						& 342.47073  & 0.1780 & 0.3088 &0.2230 & 0.2039& 0.2055  \\
H$_2$S						& 398.69486  & 0.1808 & 0.3386 &0.2406 & 0.2203& 0.2214  \\
HCl						& 460.08973  & 0.1745 & 0.3207 &0.2305 & 0.2133& 0.2164  \\
Li$_2$						& 14.87012   & 0.0337 & 0.0666 &0.0400 & 0.0289& 0.0289  \\
LiF						& 106.94528  & 0.2172 & 0.2958 &0.2089 & 0.1879& 0.1917  \\
C$_2$H$_2$					& 76.82493   & 0.2957 & 0.3984 &0.2876 & 0.2601& 0.2622  \\
C$_2$H$_4$					& 78.04004   & 0.3255 & 0.4457 &0.2970 & 0.2987& 0.3024  \\
C$_2$H$_6$					& 79.23511   & 0.3582 & 0.4599 &0.3335 & 0.3252& 0.3409  \\
CN radical					& 92.19910   & 0.2936 & 0.4327 &0.2708 & 0.2475& 0.2546  \\
HCN						& 92.88003   & 0.3202 & 0.4330 &0.3108 & 0.2815& 0.2859  \\
CO						& 112.74650  & 0.3185 & 0.4563 &0.3270 & 0.2981& 0.3072  \\
HCO  radical					& 113.25228  & 0.3345 & 0.5194 &0.2943 & 0.3274& 0.3330  \\
H$_2$CO						& 113.87497  & 0.3547 & 0.5536 &0.3787 & 0.3553& 0.3635  \\
H$_3$COH					& 115.04900  & 0.3825 & 0.5629 &0.4008 & 0.3765& 0.3886  \\
N$_2$						& 108.94735  & 0.3397 & 0.4810 &0.2511 & 0.3110& 0.3230  \\
H$_2$NNH$_2$					& 111.18640  & 0.3953 & 0.5959 &0.4199 & 0.3953& 0.4130  \\
NO radical					& 129.25550  & 0.3530 & 0.6025 &0.3104 & 0.3640& 0.3734  \\
O$_2$						& 149.60042  & 0.3962 & 0.6263 &0.3256 & 0.4008& 0.4097  \\
HOOH						& 150.78276  & 0.4229 & 0.6899 &0.4600 & 0.4418& 0.4622  \\
F$_2$						& 198.68544  & 0.4242 & 0.7306 &0.4713 & 0.4571& 0.4886  \\
CO$_2$						& 187.64687  & 0.5185 & 0.7456 &0.5225 & 0.5215& 0.5421  \\
Na$_2$						& 323.70499  & 0.0330 & 0.0927 &0.0557 & 0.0477& 0.0477  \\
Si$_2$						& 577.73197  & 0.2121 & 0.4450 &0.2683 & 0.2744& 0.2767  \\
P$_2$						& 681.46368  & 0.2825 & 0.5230 &0.3634 & 0.3461& 0.3541  \\
S$_2$						& 795.05045  & 0.3017 & 0.6109 &0.3291 & 0.4113& 0.4184  \\
Cl$_2$						& 918.96282  & 0.3209 & 0.6699 &0.4573 & 0.4616& 0.4797  \\
NaCl						& 621.43420  & 0.1669 & 0.3308 &0.2296 & 0.2115& 0.2161  \\
SiO						& 363.78893  & 0.3085 & 0.4943 &0.3319 & 0.3185& 0.3293  \\
CS						& 435.33022  & 0.2896 & 0.5030 &0.3305 & 0.3174& 0.3254  \\
SO						& 472.33136  & 0.3488 & 0.6098 &0.3688 & 0.4019& 0.4083  \\
ClO radical					& 534.25352  & 0.3372 & 0.6934 &0.4046 & 0.4371& 0.4526  \\
ClF						& 558.84497  & 0.3658 & 0.6775 &0.4563 & 0.4526& 0.4735  \\
Si$_2$H$_6$					& 581.33931  & 0.2929 & 0.4797 &0.3460 & 0.3332& 0.3622  \\
CH$_3$Cl					& 499.11859  & 0.3366 & 0.5649 &0.4021 & 0.3870& 0.3997  \\
H$_3$C-SH					& 437.72609  & 0.3449 & 0.5803 &0.4132 & 0.3957& 0.4029  \\
HOCl						& 534.87324  & 0.3717 & 0.6835 &0.4631 & 0.4559& 0.4721  \\
\hline\\[4mm]
\caption{Correlation energies $E_c$ for the G2-1 test set (the first 55 systems
of Table~\ref{tab:correl}) calculated with the cc-pVDZ basis set. All quantities are in a.u.}
\label{tab:correl2} \\
\end{longtable*}

\newpage
\section{Atomization energies}

\subsection{Atomic total energies}

\begin{longtable*}{ldddddddd}
\setlength{\tabcolsep}{0.2truecm}
Atom         &
\multicolumn{1}{c}{R(O)HF} &
\multicolumn{1}{c}{MP2} &
\multicolumn{1}{c}{B3LYP} &
\multicolumn{1}{c}{CCSD(T)}  &
\multicolumn{1}{c}{M\"{u}ller} &
\multicolumn{1}{c}{BBC3} &
\multicolumn{1}{c}{PNOF0} &
\multicolumn{1}{c}{PNOF} \\
\hline \\[-2mm]
\endhead
\multicolumn{9}{c}{\bf 6-31G$^*$ basis} \\
H  &    0.49823 &   0.49823 &   0.50027 &   0.49823 &   0.50473 &   0.50473 &   0.50212 &   0.50212 \\ 
Li &    7.43137 &   7.43186 &   7.49098 &   7.43188 &   7.47526 &   7.45242 &   7.44091 &   7.44091 \\ 
Be &   14.56694 &  14.59645 &  14.66844 &  14.61662 &  14.69850 &  14.62789 &  14.60215 &  14.60216 \\ 
B  &   24.51942 &  24.56246 &  24.65435 &  24.58340 &  24.77254 &  24.58811 &  24.58161 &  24.58201 \\ 
C  &   37.67713 &  37.73651 &  37.84628 &  37.75620 &  37.95462 &  37.78778 &  37.76187 &  37.76309 \\ 
N  &   54.38231 &  54.45945 &  54.58449 &  54.47611 &  54.57641 &  54.50031 &  54.48117 &  54.48305 \\ 
O  &   74.77897 &  74.88200 &  75.06061 &  74.89844 &  75.16162 &  74.91413 &  74.90752 &  74.90883 \\ 
F  &   99.36179 &  99.48904 &  99.71553 &  99.50049 &  99.73506 &  99.53559 &  99.50004 &  99.50278 \\ 
Na &  161.84140 & 161.84364 & 162.27988 & 161.84381 & 161.89473 & 161.85968 & 161.85578 & 161.85581 \\ 
Al &  241.85478 & 241.89500 & 242.36823 & 241.91007 & 242.11458 & 241.93738 & 241.93492 & 241.93527 \\ 
Si &  288.82915 & 288.88207 & 289.37173 & 288.89898 & 289.11896 & 288.93167 & 288.93008 & 288.93124 \\ 
P  &  340.68999 & 340.75727 & 341.25808 & 340.77373 & 340.93201 & 340.81077 & 340.80642 & 340.80842 \\ 
S  &  397.47233 & 397.56340 & 398.10500 & 397.58177 & 397.88756 & 397.61995 & 397.63510 & 397.63667 \\ 
Cl &  459.44422 & 459.56206 & 460.13626 & 459.57946 & 459.87582 & 459.63524 & 459.63100 & 459.63410 \\ 
\\
\multicolumn{9}{c}{\bf cc-pVDZ basis} \\
H  &    0.49928 &   0.49928 &   0.50126 &   0.49928 &   0.51544 &   0.50834 &   0.50855 &   0.50855 \\ 
Li &    7.43242 &   7.43341 &   7.49202 &   7.43346 &   7.48348 &   7.45963 &   7.44376 &   7.44376 \\ 
Be &   14.57236 &  14.59996 &  14.67205 &  14.61856 &  14.69657 &  14.64292 &  14.60429 &  14.60429 \\ 
B  &   24.52663 &  24.56885 &  24.66165 &  24.59153 &  24.76772 &  24.64675 &  24.58406 &  24.58445 \\ 
C  &   37.68247 &  37.74148 &  37.85290 &  37.76333 &  37.95395 &  37.81797 &  37.76365 &  37.76489 \\ 
N  &   54.38847 &  54.46585 &  54.59024 &  54.48234 &  54.58388 &  54.51316 &  54.48844 &  54.49030 \\ 
O  &   74.78758 &  74.89891 &  75.06961 &  74.91450 &  75.18168 &  75.01424 &  74.92679 &  74.92806 \\ 
F  &   99.37193 &  99.52152 &  99.72778 &  99.53291 &  99.76946 &  99.59147 &  99.53453 &  99.53720 \\ 
Na &  161.85307 & 161.85551 & 162.29148 & 161.85567 & 161.91068 & 161.88600 & 161.86980 & 161.86987 \\ 
Al &  241.87030 & 241.91125 & 242.38415 & 241.92832 & 242.12521 & 241.99711 & 241.94758 & 241.94792 \\ 
Si &  288.84662 & 288.90373 & 289.39012 & 288.92356 & 289.13801 & 288.99332 & 288.94759 & 288.94878 \\ 
P  &  340.70920 & 340.78227 & 341.27813 & 340.80065 & 340.95621 & 340.86882 & 340.82851 & 340.83080 \\ 
S  &  397.49304 & 397.59045 & 398.12672 & 397.60987 & 397.91093 & 397.73842 & 397.65771 & 397.65927 \\ 
Cl &  459.46739 & 459.59050 & 460.16014 & 459.60779 & 459.89575 & 459.71057 & 459.65314 & 459.65632 \\ 
\hline\\[4mm]
\caption{\label{atoms1} Absolute value of the atomic total energies, calculated
  both with the 6-31G$^*$ and the cc-pVDZ basis sets. These were the values used
  for the calculation of the atomization energies presented in the following tables.
  All quantities are in a.u.
} \\
\end{longtable*}

\subsection{G2/97 set, 6-31G* basis}

\begin{longtable*}{p{5.4cm}dddddd}
\setlength{\tabcolsep}{0.5truecm}
{\bf System} & 
\multicolumn{1}{c}{$-E_{\rm a}$ CCSD(T)} & 
\multicolumn{1}{c}{$-E_{\rm a}$ MP2} & 
\multicolumn{1}{c}{$-E_{\rm a}$ B3LYP} & 
\multicolumn{1}{c}{$-E_{\rm a}$ BBC3} & 
\multicolumn{1}{c}{$-E_{\rm a}$ PNOF0}& 
\multicolumn{1}{c}{$-E_{\rm a}$ PNOF} \\ 
\hline \\[-2mm]
\endhead
LiH						& 0.0734 & 0.0664 & 0.0906 & 0.0503 & 0.0539 & 0.0539 \\
BeH						& 0.0660 & 0.0767 & 0.0923 & 0.0617 & 0.0807 & 0.0807 \\
CH radical					& 0.1132 & 0.1077 & 0.1325 & 0.0898 & 0.1043 & 0.1046 \\
CH$_2$ ($^3$B$_1$)				& 0.2740 & 0.2745 & 0.3031 & 0.2417 & 0.2623 & 0.2631 \\
CH$_2$ ($^1$A$_1$)				& 0.2494 & 0.2410 & 0.2813 & 0.2239 & 0.2323 & 0.2327 \\
CH$_3$ radical					& 0.4444 & 0.4418 & 0.4912 & 0.3994 & 0.4214 & 0.4207 \\
CH$_4$						& 0.6112 & 0.6076 & 0.6710 & 0.5520 & 0.5790 & 0.5737 \\
NH						& 0.1075 & 0.1037 & 0.1346 & 0.0881 & 0.0985 & 0.0985 \\
NH$_2$ radical					& 0.2417 & 0.2378 & 0.2875 & 0.2143 & 0.2224 & 0.2225 \\
NH$_3$						& 0.4042 & 0.4032 & 0.4626 & 0.3771 & 0.3764 & 0.3774 \\
OH radical					& 0.1424 & 0.1430 & 0.1626 & 0.1370 & 0.1276 & 0.1275 \\
H$_2$O						& 0.3151 & 0.3208 & 0.3478 & 0.2993 & 0.2825 & 0.2842 \\
HF						& 0.1914 & 0.1969 & 0.2044 & 0.1541 & 0.1690 & 0.1687 \\
SiH$_2$ ($^1$A$_1$)				& 0.2080 & 0.1987 & 0.2407 & 0.2131 & 0.2043 & 0.2036 \\
SiH$_2$ ($^3$B$_1$)				& 0.1811 & 0.1776 & 0.2087 & 0.1789 & 0.1779 & 0.1776 \\
SiH$_3$ radical					& 0.3145 & 0.3074 & 0.3596 & 0.3054 & 0.3047 & 0.3031 \\
SiH$_4$						& 0.4520 & 0.4419 & 0.5108 & 0.4356 & 0.4439 & 0.4333 \\
PH$_2$ radical					& 0.2014 & 0.1921 & 0.2455 & 0.2001 & 0.2009 & 0.2014 \\
PH$_3$						& 0.3223 & 0.3103 & 0.3812 & 0.3264 & 0.3191 & 0.3198 \\
H$_2$S						& 0.2448 & 0.2388 & 0.2798 & 0.2647 & 0.2330 & 0.2334 \\
HCl						& 0.1429 & 0.1419 & 0.1591 & 0.1461 & 0.1348 & 0.1349 \\
Li$_2$						& 0.0342 & 0.0231 & 0.0320 &-0.0040 & 0.0081 & 0.0081 \\
LiF						& 0.1994 & 0.2086 & 0.2110 & 0.1363 & 0.1631 & 0.1613 \\
C$_2$H$_2$					& 0.5942 & 0.6067 & 0.6323 & 0.4931 & 0.5441 & 0.5445 \\
C$_2$H$_4$					& 0.8258 & 0.8283 & 0.8937 & 0.7225 & 0.7864 & 0.7858 \\
C$_2$H$_6$					& 1.0418 & 1.0416 & 1.1361 & 0.9364 & 1.0078 & 0.9959 \\
CN radical					& 0.2541 & 0.2460 & 0.2776 & 0.1758 & 0.2021 & 0.2013 \\
HCN						& 0.4555 & 0.4728 & 0.4908 & 0.3557 & 0.4058 & 0.4055 \\
CO						& 0.3902 & 0.4097 & 0.4022 & 0.3551 & 0.3649 & 0.3582 \\
HCO  radical					& 0.4073 & 0.4236 & 0.4429 & 0.3189 & 0.3893 & 0.3863 \\
H$_2$CO						& 0.5465 & 0.5600 & 0.5927 & 0.5164 & 0.5364 & 0.5312 \\
H$_3$COH					& 0.7340 & 0.7419 & 0.8064 & 0.6916 & 0.7182 & 0.7105 \\
N$_2$						& 0.3219 & 0.3427 & 0.3535 & 0.1841 & 0.2899 & 0.2817 \\
H$_2$NNH$_2$					& 0.5897 & 0.5926 & 0.6864 & 0.5511 & 0.5891 & 0.5741 \\
NO radical					& 0.2091 & 0.2230 & 0.2425 & 0.1260 & 0.2115 & 0.2031 \\
O$_2$						& 0.1687 & 0.1903 & 0.1975 & 0.1193 & 0.1580 & 0.1520 \\
HOOH						& 0.3622 & 0.3745 & 0.4114 & 0.3621 & 0.3808 & 0.3625 \\
F$_2$						& 0.0537 & 0.0607 & 0.0669 & 0.0321 & 0.1144 & 0.0878 \\
CO$_2$						& 0.5785 & 0.6179 & 0.6131 & 0.5263 & 0.5766 & 0.5599 \\
Na$_2$						& 0.0252 & 0.0167 & 0.0269 & 0.0062 & 0.0062 & 0.0063 \\
Si$_2$						& 0.0985 & 0.0965 & 0.1060 & 0.1007 & 0.1105 & 0.1106 \\
P$_2$						& 0.1459 & 0.1502 & 0.1723 & 0.1636 & 0.1585 & 0.1539 \\
S$_2$						& 0.1305 & 0.1360 & 0.1529 & 0.1293 & 0.1438 & 0.1384 \\
Cl$_2$						& 0.0649 & 0.0671 & 0.0770 & 0.0867 & 0.1147 & 0.1050 \\
NaCl						& 0.1376 & 0.1406 & 0.1442 & 0.1274 & 0.1194 & 0.1182 \\
SiO						& 0.2763 & 0.2953 & 0.2873 & 0.2467 & 0.2576 & 0.2491 \\
CS						& 0.2503 & 0.2578 & 0.2580 & 0.2299 & 0.2298 & 0.2268 \\
SO						& 0.1662 & 0.1813 & 0.1869 & 0.1381 & 0.1625 & 0.1606 \\
ClO radical					& 0.0762 & 0.0746 & 0.0954 & 0.0644 & 0.1292 & 0.1192 \\
ClF						& 0.0813 & 0.0882 & 0.0907 & 0.0881 & 0.1373 & 0.1226 \\
Si$_2$H$_6$					& 0.7461 & 0.7316 & 0.8374 & 0.7283 & 0.7585 & 0.7344 \\
CH$_3$Cl					& 0.5731 & 0.5758 & 0.6249 & 0.5455 & 0.5756 & 0.5690 \\
H$_3$C-SH					& 0.6768 & 0.6750 & 0.7457 & 0.6658 & 0.6722 & 0.6707 \\
HOCl						& 0.2189 & 0.2272 & 0.2463 & 0.2319 & 0.2571 & 0.2472 \\
SO$_2$						& 0.3354 & 0.3727 & 0.3608 & 0.3346 & 0.3926 & 0.3599 \\
BF$_3$						& 0.7225 & 0.7620 & 0.7522 & 0.6161 & 0.7816 & 0.7361 \\
BCl$_3$						& 0.4866 & 0.5110 & 0.4991 & 0.4942 & 0.5827 & 0.5504 \\
AlF$_3$						& 0.6391 & 0.6776 & 0.6554 & 0.5090 & 0.6942 & 0.6251 \\
AlCl$_3$ 					& 0.4471 & 0.4672 & 0.4559 & 0.4483 & 0.5403 & 0.4971 \\
CF$_4$						& 0.7274 & 0.7696 & 0.7679 & 0.5856 & 0.8771 & 0.7745 \\
CCl$_4$						& 0.4429 & 0.4681 & 0.4615 & 0.4392 & 0.6112 & 0.5384 \\
O=C=S						& 0.4920 & 0.5225 & 0.5232 & 0.4872 & 0.5100 & 0.5003 \\
CS$_2$						& 0.4062 & 0.4284 & 0.4325 & 0.4245 & 0.4388 & 0.4312 \\
COF$_2$						& 0.6291 & 0.6686 & 0.6698 & 0.5664 & 0.7106 & 0.6472 \\
SiF$_4$						& 0.8561 & 0.9024 & 0.8745 & 0.6994 & 0.9733 & 0.8576 \\
SiCl$_4$					& 0.5371 & 0.5613 & 0.5551 & 0.5540 & 0.7045 & 0.6207 \\
N$_2$O  					& 0.3693 & 0.4128 & 0.4279 & 0.3356 & 0.3788 & 0.3704 \\
ClNO						& 0.2587 & 0.2799 & 0.3032 & 0.2783 & 0.3403 & 0.3278 \\
NF$_3$						& 0.2872 & 0.3100 & 0.3402 & 0.2337 & 0.4537 & 0.3852 \\
PF$_3$						& 0.5195 & 0.5482 & 0.5516 & 0.4616 & 0.6442 & 0.5912 \\
O$_3$						& 0.1886 & 0.2308 & 0.2221 & 0.0821 & 0.2877 & 0.2347 \\
F$_2$O						& 0.1248 & 0.1397 & 0.1578 & 0.1300 & 0.2577 & 0.2150 \\
ClF$_3$						& 0.1473 & 0.1726 & 0.1820 & 0.1073 & 0.3850 & 0.2586 \\
C$_2$F$_4$					& 0.8836 & 0.9315 & 0.9448 & 0.7371 & 1.0496 & 0.9821 \\
C$_2$Cl$_4$					& 0.6838 & 0.7185 & 0.7113 & 0.6418 & 0.8468 & 0.6510 \\
CF$_3$CN  					& 0.9615 & 1.0175 & 1.0219 & 0.7525 & 1.0775 & 0.9535 \\
CH$_3$CCH (propyne)				& 1.0368 & 1.0538 & 1.1131 & 0.9274 & 0.9990 & 0.9911 \\
CH$_2$=C=CH$_2$ (allene)			& 1.0357 & 1.0461 & 1.1176 & 0.8703 & 1.0106 & 1.0020 \\
C$_3$H$_4$ (cyclopropene)			& 1.0010 & 1.0171 & 1.0790 & 0.8963 & 0.9706 & 0.9679 \\
CH$_3$CH=CH$_2$ (propylene)			& 1.2639 & 1.2707 & 1.3670 & 1.1229 & 1.2305 & 1.2246 \\
C$_3$H$_6$ (cyclopropane)			& 1.2524 & 1.2639 & 1.3546 & 1.1305 & 1.2163 & 1.2047 \\
C$_3$H$_8$ (propane)				& 1.4756 & 1.4790 & 1.6030 & 1.3202 & 1.4432 & 1.4196 \\
CH$_2$CHCHCH$_2$ (Trans-1,3-butadiene)		& 1.4918 & 1.5063 & 1.6053 & 1.3392 & 1.4637 & 1.4565 \\
C$_4$H$_6$ (2-butyne)				& 1.4783 & 1.4997 & 1.5919 & 1.3268 & 1.4433 & 1.4282 \\
C$_4$H$_6$ (methylene cyclopropane)		& 1.4618 & 1.4806 & 1.5758 & 1.3147 & 1.4295 & 1.4118 \\
C$_4$H$_6$ (bicyclobutane)		 	& 1.4497 & 1.4741 & 1.5611 & 1.3027 & 1.4185 & 1.4136 \\
C$_4$H$_6$ (cyclobutene)			& 1.4750 & 1.4939 & 1.5864 & 1.2915 & 1.4357 & 1.4267 \\
C$_4$H$_8$ (cyclobutane)			& 1.6908 & 1.7052 & 1.8254 & 1.4959 & 1.6543 & 1.6243 \\
C$_4$H$_8$ (isobutene)				& 1.7032 & 1.7145 & 1.8398 & 1.5453 & 1.6624 & 1.6500 \\
C$_4$H$_{10}$ (trans-butane)			& 1.9096 & 1.9166 & 2.0699 & 1.6938 & 1.8711 & 1.8329 \\
C$_4$H$_{10}$ (isobutane)			& 1.9120 & 1.9194 & 2.0707 & 1.6941 & 1.8696 & 1.8317 \\
C$_5$H$_8$ (spiropentane)			& 1.8931 & 1.9208 & 2.0378 & 1.6535 & 1.8477 & 1.8222 \\
C$_6$H$_6$ (benzene)				& 2.0333 & 2.0788 & 2.1694 & 1.7546 & 1.9548 & 1.9386 \\
H$_2$CF$_2$					& 0.6438 & 0.6623 & 0.6952 & 0.5387 & 0.6894 & 0.6425 \\
HCF$_3$						& 0.6867 & 0.7171 & 0.7328 & 0.5510 & 0.7845 & 0.7201 \\
H$_2$CCl$_2$					& 0.5340 & 0.5436 & 0.5765 & 0.5177 & 0.5949 & 0.5713 \\
HCCl$_3$					& 0.4914 & 0.5085 & 0.5230 & 0.4816 & 0.6218 & 0.5682 \\
H$_3$C-NH$_2$ (methylamine)			& 0.8254 & 0.8273 & 0.9210 & 0.7718 & 0.8075 & 0.8018 \\
CH$_3$-CN (acetonitrile)			& 0.9027 & 0.9241 & 0.9763 & 0.8014 & 0.8735 & 0.8567 \\
CH$_3$-NO$_2$ (nitromethane)			& 0.8544 & 0.8907 & 0.9559 & 0.7945 & 0.9274 & 0.8736 \\
CH$_3$ONO (methylnitrite)			& 0.8560 & 0.8845 & 0.9546 & 0.8111 & 0.9201 & 0.8842 \\
CH$_3$-SiH$_3$ (methylsilane)			& 0.8983 & 0.8924 & 0.9911 & 0.8378 & 0.8858 & 0.8631 \\
HCOOH (formic acid)				& 0.7269 & 0.7549 & 0.7873 & 0.6922 & 0.7344 & 0.7067 \\
HCOOCH$_3$ (methyl formate)			& 1.1512 & 1.1817 & 1.2479 & 1.0761 & 1.1808 & 1.1396 \\
CH$3$CONH$_2$ (acetamide)			& 1.2534 & 1.2794 & 1.3729 & 1.1655 & 1.2719 & 1.2409 \\
C$_2$H$_4$NH (aziridine)			& 1.0366 & 1.0495 & 1.1409 & 0.9533 & 1.0209 & 1.0133 \\
NCCN (cyanogen)   				& 0.7366 & 0.7827 & 0.7912 & 0.5998 & 0.7153 & 0.6628 \\
(CH$_3$)$_2$NH (dimethylamine) 			& 1.2528 & 1.2580 & 1.3838 & 1.1686 & 1.2400 & 1.2318 \\
CH$_3$CH$_2$NH$_2$ (trans-ethylamine)		& 1.2627 & 1.2682 & 1.3916 & 1.1716 & 1.2450 & 1.2338 \\
H$_2$CCO (ketene)				& 0.7829 & 0.8086 & 0.8446 & 0.7370 & 0.7906 & 0.7857 \\
C$_2$H$_4$O (oxirane)				& 0.9490 & 0.9678 & 1.0319 & 0.8869 & 0.9401 & 0.9162 \\
CH$_3$CHO (acetaldehyde)			& 0.9934 & 1.0110 & 1.0756 & 0.9308 & 0.9912 & 0.9787 \\
OCHCHO (glyoxal)				& 0.9350 & 0.9703 & 1.0037 & 0.8655 & 1.0013 & 0.8882 \\
CH$_3$CH$_2$OH (ethanol)			& 1.1730 & 1.1845 & 1.2788 & 1.0912 & 1.1663 & 1.1291 \\
CH$_3$-O-CH$_3$	(dimethylether)			& 1.1606 & 1.1711 & 1.2701 & 1.0877 & 1.1555 & 1.1426 \\
C$_2$H$_4$S (thiooxirane)			& 0.9088 & 0.9203 & 0.9845 & 0.8786 & 0.9105 & 0.9060 \\
(CH$_3$)$_2$SO (dimethylsulfoxide)		& 1.2119 & 1.2285 & 1.3264 & 1.1704 & 1.2474 & 1.2287 \\
C$_2$H$_5$SH (ethanethiol)			& 1.1117 & 1.1139 & 1.2134 & 1.0639 & 1.1154 & 1.1005 \\
CH$_3$-S-CH$_3$ (dimethyl sulfide)		& 1.1128 & 1.1155 & 1.2142 & 1.0724 & 1.1183 & 1.1113 \\
H$_2$C=CHF (vinylfluoride)			& 0.8442 & 0.8584 & 0.9106 & 0.7614 & 0.8503 & 0.8424 \\
C$_2$H$_5$Cl (ethylchloride)			& 1.0106 & 1.0172 & 1.0957 & 0.9392 & 1.0207 & 0.9999 \\
CH$_2$=CHCl (vinylchloride)			& 0.7962 & 0.8063 & 0.8555 & 0.7462 & 0.8101 & 0.8034 \\
H$_2$C=CHCN (acrylonitrile)			& 1.1233 & 1.1524 & 1.2067 & 0.9837 & 1.0973 & 1.0850 \\
CH$_3$-CO-CH$_3$ (acetone)			& 1.4383 & 1.4600 & 1.5543 & 1.3311 & 1.4361 & 1.4175 \\
CH$_3$COOH (acetic acid)			& 1.1723 & 1.2040 & 1.2668 & 1.0962 & 1.1941 & 1.1582 \\
CH$_3$COF (acetyl fluoride)			& 1.0438 & 1.0746 & 1.1226 & 0.9542 & 1.0853 & 1.0421 \\
CH$_3$COCl (acetyl chloride)			& 0.9805 & 1.0069 & 1.0551 & 0.9231 & 1.0324 & 0.9718 \\
CH$_3$CH$_2$CH$_2$Cl (propyl chloride)		& 1.4449 & 1.4552 & 1.5627 & 1.3138 & 1.4524 & 1.4173 \\
(CH$_3$)$_2$CHOH (isopropanol)			& 1.6138 & 1.6292 & 1.7515 & 1.4757 & 1.6025 & 1.5680 \\
C$_2$H$_5$-O-CH$_3$ (methyl-ethyl ether)	& 1.5999 & 1.6141 & 1.7425 & 1.4725 & 1.5968 & 1.5432 \\
(CH$_3$)$_3$N (trimethylamine)			& 1.6843 & 1.6934 & 1.8483 & 1.5525 & 1.6648 & 1.6522 \\
C$_4$H$_4$O (furan)				& 1.4699 & 1.5118 & 1.5736 & 1.3299 & 1.4535 & 1.4468 \\
C$_4$H$_4$S (thiophene)				& 1.4184 & 1.4536 & 1.5110 & 1.2918 & 1.4085 & 1.3955 \\
C$_4$H$_4$NH (pyrrole)				& 1.5686 & 1.6075 & 1.6948 & 1.4081 & 1.5450 & 1.5409 \\
C$_5$H$_5$N (pyridine)				& 1.8301 & 1.8775 & 1.9677 & 1.5770 & 1.7827 & 1.7476 \\
H$_2$						& 0.1552 & 0.1477 & 0.1749 & 0.1425 & 0.1375 & 0.1375 \\
SH radical					& 0.1142 & 0.1105 & 0.1347 & 0.1218 & 0.1088 & 0.1090 \\
CCH radical					& 0.3854 & 0.3822 & 0.4097 & 0.3112 & 0.3388 & 0.3392 \\
C$_2$H$_3$ radical ($^2$A$'$)			& 0.6494 & 0.6455 & 0.7071 & 0.5515 & 0.6150 & 0.6155 \\
CH$_3$CO radical ($^2$A$'$)			& 0.8534 & 0.8729 & 0.9257 & 0.7300 & 0.8509 & 0.8337 \\
H$_2$COH radical ($^2$A)			& 0.5814 & 0.5902 & 0.6443 & 0.5366 & 0.5685 & 0.5637 \\
CH$_3$O radical (Cs)				& 0.5788 & 0.5799 & 0.6424 & 0.5304 & 0.5786 & 0.5709 \\
CH$_3$CH$_2$O radical ($^2$A$''$)		& 1.0164 & 1.0209 & 1.1140 & 0.9257 & 1.0244 & 1.0043 \\
CH$_3$S radical ($^2$A$'$)			& 0.5512 & 0.5514 & 0.6075 & 0.5199 & 0.5531 & 0.5502 \\
C$_2$H$_5$ radical ($^2$A$'$)			& 0.8802 & 0.8805 & 0.9638 & 0.7958 & 0.8508 & 0.8471 \\
(CH$_3$)$_2$CH radical ($^2$A$'$)		& 1.3186 & 1.3221 & 1.4372 & 1.1922 & 1.2886 & 1.2769 \\
(CH$_3$)$_3$C (t-butyl radical)			& 1.7585 & 1.7657 & 1.9102 & 1.5768 & 1.7166 & 1.6987 \\
NO$_2$ radical					& 0.3098 & 0.3452 & 0.3660 & 0.1830 & 0.3533 & 0.3393 \\
\hline\\[4mm]
\caption{Atomization energies $E_{\rm a}$ for the full G2/97 test set calculated with
the 6-31G* basis. 
All quantities are in a.u.} \label{tab:atomiz_all} \\
\end{longtable*}

\subsection{G2-1 set, cc-pVDZ basis}

\begin{longtable*}{p{5.4cm}dddddd}
\setlength{\tabcolsep}{0.5truecm}
{\bf System} & 
\multicolumn{1}{c}{$-E_{\rm a}$ CCSD(T)} & 
\multicolumn{1}{c}{$-E_{\rm a}$ MP2} & 
\multicolumn{1}{c}{$-E_{\rm a}$ B3LYP} & 
\multicolumn{1}{c}{$-E_{\rm a}$ BBC3} & 
\multicolumn{1}{c}{$-E_{\rm a}$ PNOF0}& 
\multicolumn{1}{c}{$-E_{\rm a}$ PNOF} \\ 
\hline \\[-2mm]
\endhead
LiH						& 0.0833 & 0.0749 & 0.0908 & 0.0512  & 0.0554 & 0.0554 \\
BeH						& 0.0726 & 0.0802 & 0.0892 & 0.0612  & 0.0772 & 0.0772 \\
CH radical					& 0.1211 & 0.1155 & 0.1315 & 0.0890  & 0.1056 & 0.1053 \\
CH$_2$ ($^3$B$_1$)				& 0.2814 & 0.2815 & 0.2977 & 0.2391  & 0.2597 & 0.2597 \\
CH$_2$ ($^1$A$_1$)				& 0.2621 & 0.2532 & 0.2781 & 0.2215  & 0.2340 & 0.2336 \\
CH$_3$ radical					& 0.4570 & 0.4536 & 0.4821 & 0.3925  & 0.4152 & 0.4160 \\
CH$_4$						& 0.6294 & 0.6242 & 0.6582 & 0.5410  & 0.5647 & 0.5706 \\
NH						& 0.1146 & 0.1101 & 0.1330 & 0.0952  & 0.0972 & 0.0973 \\
NH$_2$ radical					& 0.2571 & 0.2518 & 0.2858 & 0.2207  & 0.2202 & 0.2206 \\
NH$_3$						& 0.4277 & 0.4244 & 0.4609 & 0.3825  & 0.3745 & 0.3735 \\
OH radical					& 0.1509 & 0.1503 & 0.1625 & 0.1442  & 0.1258 & 0.1259 \\
H$_2$O						& 0.3339 & 0.3372 & 0.3497 & 0.3069  & 0.2825 & 0.2827 \\
HF						& 0.2018 & 0.2066 & 0.2079 & 0.1621  & 0.1690 & 0.1693 \\
SiH$_2$ ($^1$A$_1$)				& 0.2248 & 0.2141 & 0.2385 & 0.2122  & 0.2031 & 0.2038 \\
SiH$_2$ ($^3$B$_1$)				& 0.1947 & 0.1907 & 0.2064 & 0.1773  & 0.1758 & 0.1764 \\
SiH$_3$ radical					& 0.3365 & 0.3276 & 0.3552 & 0.3019  & 0.3000 & 0.3020 \\
SiH$_4$						& 0.4824 & 0.4689 & 0.5039 & 0.4312  & 0.4294 & 0.4409 \\
PH$_2$ radical					& 0.2199 & 0.2097 & 0.2448 & 0.1919  & 0.2015 & 0.2005 \\
PH$_3$						& 0.3502 & 0.3367 & 0.3789 & 0.3221  & 0.3205 & 0.3197 \\
H$_2$S						& 0.2672 & 0.2615 & 0.2816 & 0.2667  & 0.2403 & 0.2399 \\
HCl						& 0.1571 & 0.1564 & 0.1622 & 0.1559  & 0.1414 & 0.1413 \\
Li$_2$						& 0.0369 & 0.0255 & 0.0326 & -0.0013 & 0.0115 & 0.0115 \\
LiF						& 0.1961 & 0.2050 & 0.2122 & 0.1377  & 0.1549 & 0.1560 \\
C$_2$H$_2$					& 0.5954 & 0.6113 & 0.6265 & 0.5109  & 0.5406 & 0.5403 \\
C$_2$H$_4$					& 0.8418 & 0.8459 & 0.8818 & 0.7112  & 0.7772 & 0.7784 \\
C$_2$H$_6$					& 1.0709 & 1.0697 & 1.1169 & 0.9185  & 0.9817 & 0.9950 \\
CN radical					& 0.2471 & 0.2414 & 0.2744 & 0.1840  & 0.1945 & 0.1985 \\
HCN						& 0.4553 & 0.4738 & 0.4870 & 0.3928  & 0.4009 & 0.4022 \\
CO						& 0.3872 & 0.4071 & 0.4002 & 0.3628  & 0.3541 & 0.3608 \\
HCO  radical					& 0.4097 & 0.4257 & 0.4384 & 0.3239  & 0.3807 & 0.3838 \\
H$_2$CO						& 0.5533 & 0.5656 & 0.5839 & 0.5187  & 0.5227 & 0.5285 \\
H$_3$COH					& 0.7566 & 0.7619 & 0.7966 & 0.6906  & 0.7009 & 0.7104 \\
N$_2$						& 0.3224 & 0.3427 & 0.3536 & 0.2041  & 0.2815 & 0.2898 \\
H$_2$NNH$_2$					& 0.6199 & 0.6196 & 0.6828 & 0.5633  & 0.5706 & 0.5846 \\
NO radical					& 0.2117 & 0.2242 & 0.2432 & 0.1467  & 0.2043 & 0.2106 \\
O$_2$						& 0.1676 & 0.1875 & 0.1962 & 0.0822  & 0.1476 & 0.1540 \\
HOOH						& 0.3782 & 0.3866 & 0.4113 & 0.3746  & 0.3538 & 0.3717 \\
F$_2$						& 0.0438 & 0.0493 & 0.0602 & 0.0354  & 0.0735 & 0.0996 \\
CO$_2$						& 0.5731 & 0.6123 & 0.6084 & 0.5368  & 0.5511 & 0.5680 \\
Na$_2$						& 0.0266 & 0.0174 & 0.0270 & 0.0104  & 0.0131 & 0.0130 \\
Si$_2$						& 0.0969 & 0.0979 & 0.1069 & 0.1050  & 0.1112 & 0.1111 \\
P$_2$						& 0.1449 & 0.1511 & 0.1739 & 0.1568  & 0.1528 & 0.1562 \\
S$_2$						& 0.1324 & 0.1406 & 0.1554 & 0.0905  & 0.1463 & 0.1504 \\
Cl$_2$						& 0.0681 & 0.0723 & 0.0801 & 0.1158  & 0.1181 & 0.1299 \\
NaCl						& 0.1376 & 0.1418 & 0.1445 & 0.1365  & 0.1228 & 0.1241 \\
SiO						& 0.2594 & 0.2784 & 0.2772 & 0.2513  & 0.2331 & 0.2414 \\
CS						& 0.2467 & 0.2574 & 0.2576 & 0.2275  & 0.2263 & 0.2314 \\
SO						& 0.1558 & 0.1703 & 0.1820 & 0.1337  & 0.1488 & 0.1523 \\
ClO radical					& 0.0685 & 0.0672 & 0.0924 & 0.0840  & 0.1107 & 0.1217 \\
ClF						& 0.0701 & 0.0767 & 0.0862 & 0.0884  & 0.1099 & 0.1250 \\
Si$_2$H$_6$					& 0.7894 & 0.7715 & 0.8259 & 0.7172  & 0.7260 & 0.7527 \\
CH$_3$Cl					& 0.5862 & 0.5893 & 0.6137 & 0.5434  & 0.5632 & 0.5714 \\
H$_3$C-SH					& 0.7007 & 0.6995 & 0.7358 & 0.6576  & 0.6663 & 0.6706 \\
HOCl						& 0.2234 & 0.2304 & 0.2459 & 0.2501  & 0.2407 & 0.2524 \\
\hline\\[4mm]
\caption{Atomization energies $E_{\rm a}$ for the G2-1 test set (the first 55 systems
of Table~\ref{tab:atomiz_all}) calculated with the cc-pVDZ basis set.
All quantities are in a.u.} \label{tab:atomiz_all1} \\
\end{longtable*}
\end{widetext}

\end{document}